\documentclass[iop, numberedappendix, appendixfloats]{emulateapj}
\usepackage{graphicx,natbib,mathtools,multirow}
\usepackage{booktabs}
\usepackage{amssymb}
\usepackage{placeins}

\shorttitle{The Minimum of Solar Cycle 23:  As Deep as It Could Be?}
\shortauthors{Mu\~noz-Jaramillo et al.}

\begin{document}


\title{The Minimum of Solar Cycle 23:  As Deep as It Could Be?}

\author{Andr\'es Mu\~noz-Jaramillo\altaffilmark{1, 2, 3}, Ryan R. Senkpeil\altaffilmark{4}, Dana W.\ Longcope\altaffilmark{1}, Andrey G.\ Tlatov\altaffilmark{5}, Alexei A.\ Pevtsov\altaffilmark{6}, Laura A.\ Balmaceda\altaffilmark{7,8}, Edward E.\ DeLuca\altaffilmark{9}, and Petrus C.\ H.\ Martens\altaffilmark{10}}
\affil{$^1$ Department of Physics, Montana State University, Bozeman, MT 59717, USA; \href{mailto:munoz@solar.physics.montana.edu}{munoz@solar.physics.montana.edu}}
\affil{$^2$ Space Sciences Laboratory, University of California, Berkeley, CA 94720, USA}
\affil{$^3$ W.W.\ Hansen Experimental Physics Laboratory, Stanford University, Stanford, CA 94305, USA}
\affil{$^4$ Department of Physics, Purdue University, West Lafayette, IN 47907, USA}
\affil{$^5$ Kislovodsk Mountain Astronomical Station of the Pulkovo Observatory, Kislovodsk 357700, Russia}
\affil{$^6$ National Solar Observatory, Sunspot, NM 88349, USA}
\affil{$^7$ Institute for Astronomical, Terrestrial and Space Sciences (ICATE-CONICET), San Juan, Argentina}
\affil{$^8$ National Institute for Space Research (INPE), S\~ao Jos\'e dos Campos, Brazil}
\affil{$^9$ Harvard-Smithsonian Center for Astrophysics, Cambridge, MA 02138, USA}
\affil{$^{10}$ Department of Physics and Astronomy, Georgia State University, Atlanta, GA 30303, USA}

\begin{abstract}
In this work we introduce a new way of binning sunspot group data with the purpose of better understanding the impact of the solar cycle on sunspot properties and how this defined the characteristics of the extended minimum of cycle 23.  Our approach assumes that the statistical properties of sunspots are completely determined by the strength of the underlying large-scale field and have no additional time dependencies.  We use the amplitude of the cycle at any given moment (something we refer to as activity level) as a proxy for the strength of this deep-seated magnetic field.


We find that the sunspot size distribution is composed of two populations: one population of groups and active regions and a second population of pores and ephemeral regions.  When fits are performed at periods of different activity level, only the statistical properties of the former population, the active regions, is found to vary.

Finally, we study the relative contribution of each component (small-scale versus large-scale) to solar magnetism.  We find that when hemispheres are treated separately, almost every one of the past 12 solar minima reaches a point where the main contribution to magnetism comes from the small-scale component.  However, due to asymmetries in cycle phase, this state is very rarely reached by both hemispheres at the same time.  From this we infer that even though each hemisphere did reach the magnetic baseline, from a heliospheric point of view the minimum of cycle 23 was not as deep as it could have been.
\end{abstract}


\section{Introduction}

The solar magnetic cycle is a process that takes the Sun through subsequent periods of high (maximum) and low (minimum) activity.  Since the pioneering work of \cite{parker1955a}, there has been a continuous effort to understand the mechanisms that keep it going and define its properties \citep[see a review by][]{charbonneau2010}.  Most of this effort has focused on understanding the periods of highest activity (solar maximum). However, this focus shifted after the arrival of the unexpectedly deep minimum of solar cycle 23, in which record lows were measured across the board in solar activity indices and solar wind properties.

One of the most direct proxies of solar activity, and the one most commonly used to understand cycle variability, is the presence of sunspots on the photosphere. One of its main advantages is that reliable sunspot records exist that span more than a century.  Sunspot groups are associated with bipolar magnetic regions (BMRs), that are believed to originate from an underlying large-scale toroidal field that is critical for the evolution of the solar cycle \citep[see a review by][]{fan2009}.  Furthermore, thanks to the systematic orientation and tilt of BMRs, their emergence and decay of seem to be the primary mechanism that regenerates the poloidal field from which the solar cycle can start again \citep{dasiespuig-etal2010,cameron-etal2010,munoz-etal2013a}, a process known as the Babcock-Leighton mechanism \citep[BL;][]{babcock1961,leighton1964}.

One of the main questions that has been in the forefront of our minds since the arrival of the deep minimum of cycle 23, is: Is there a baseline for solar magnetism beyond which activity indices and solar wind properties cannot be any weaker? \citep{svalgaard-cliver2007,schrijver-etal2011,cliver2012,wang-sheeley2013}.

Both the length of the sunspot record and the direct connection between sunspots and the solar cycle make it an ideal proxy with which to probe this question.  However, distilling information out of sunspot (and BMR) properties is non-trivial due to the large variability observed in their properties (area, flux, and tilt and time, latitude, and longitude of emergence).  The main objective of this paper is to introduce a new technique for studying the cycle dependence of sunspot properties (specifically sunspot group area) and demonstrate how powerful it is for contextualizing the extended minimum of cycle 23.

In order to better explain this new approach, we begin with a brief overview of the traditional approaches for binning sunspot and BMR properties to study the solar cycle (by date, by cycle, and by cycle phase; see Section \ref{Sec_Time_bin}).  Then we discuss why these approaches are sub-optimal when the object of interest is the sunspots themselves, and we define activity level and how to use it to bin sunspot data (see Section \ref{Sec_AL_bin}).  In Section \ref{Sec_Data}, we introduce our two databases and how to combine them by taking advantage of their statistical properties followed by our statistical model and our methods for fitting the data and assessing their relative performance (see Section \ref{Sec_Math}).  In Section \ref{Sec_Dis_Free} we separate our data according to activity level and fit each activity level bin separately.  We also show how the area distribution related to small structures is independent of activity level.  In Section \ref{Sec_Dis_Con} we show how the area distribution related to BMRs is strongly dependent on activity level and re-define our statistical model to take advantage of this relationship.  In Section \ref{Sec_Model_Sel} we demonstrate quantitatively, that an activity-level-dependent statistical model is the best of all the models proposed in this paper for characterizing our data.  In Section \ref{Sec_Time_Dep}, we take advantage of this model to better understand the extended minimum of cycle 23 and put it in the context of the last 12 cycles.  In Section \ref{Sec_Dynamo} we discuss whether the existence of a composite distribution arises from different components of the dynamo (small-scale versus global), or whether an alternative interpretation of the results is necessary.  We finish with a summary and conclusions in Section \ref{Sec_Conclusions}.

\begin{figure*}[ht!]
\begin{center}
  \includegraphics[width=0.75\textwidth]{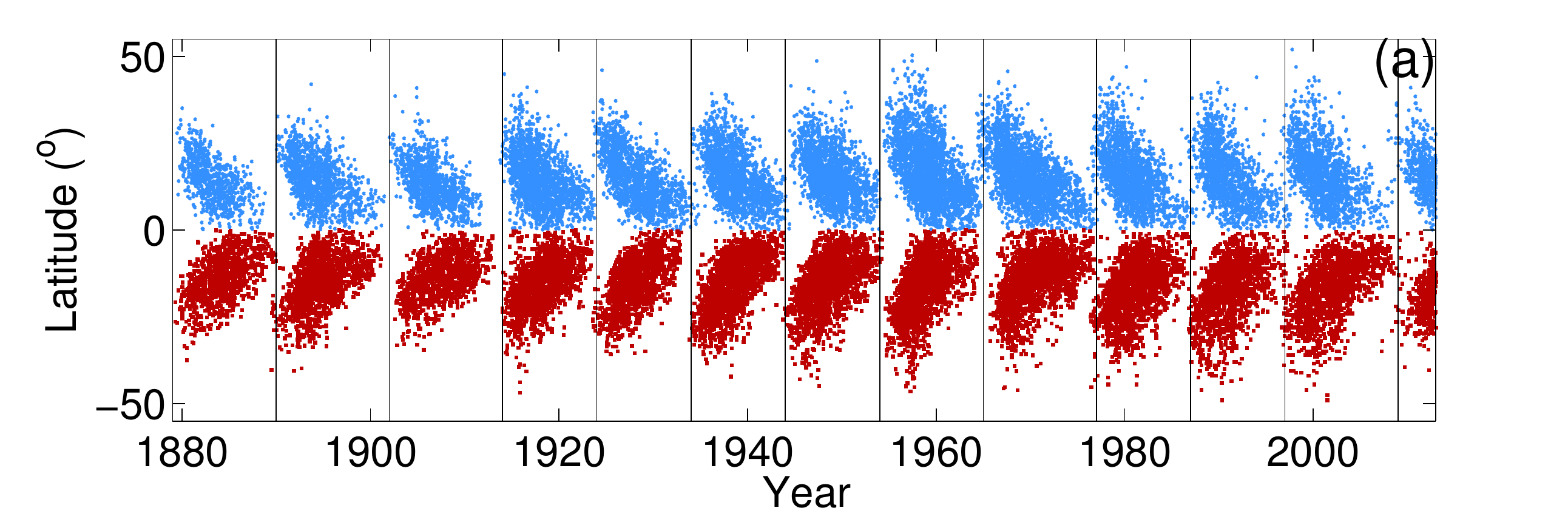}\\
  \includegraphics[width=0.75\textwidth]{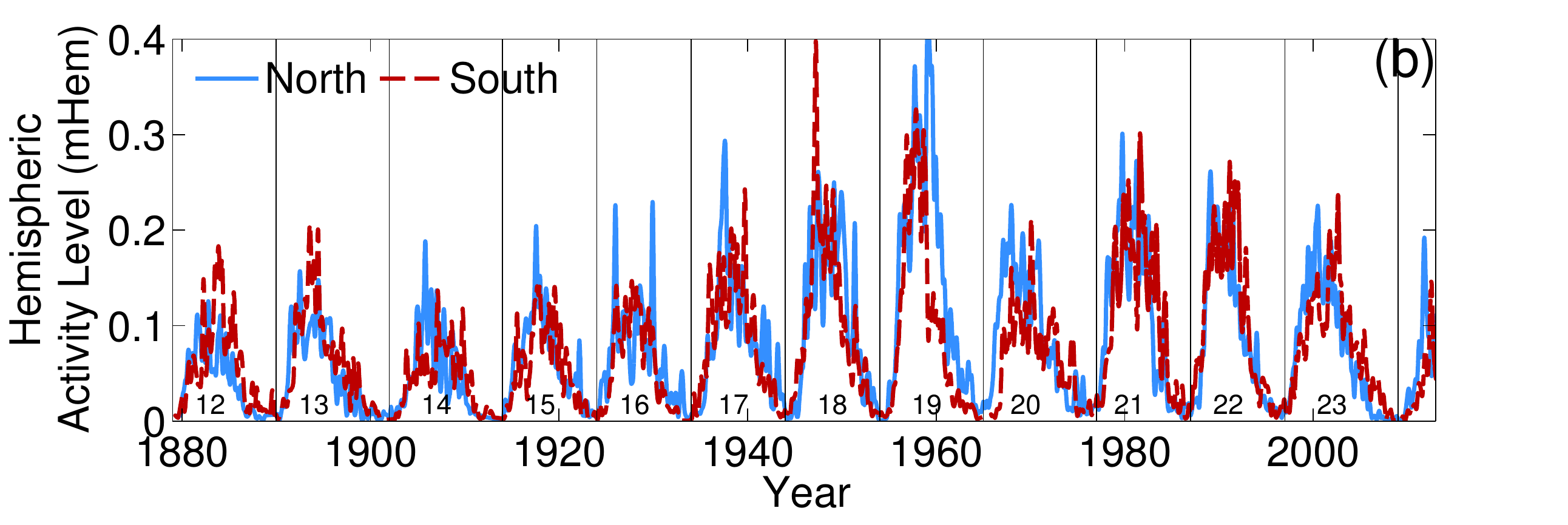}\\
  \includegraphics[width=0.75\textwidth]{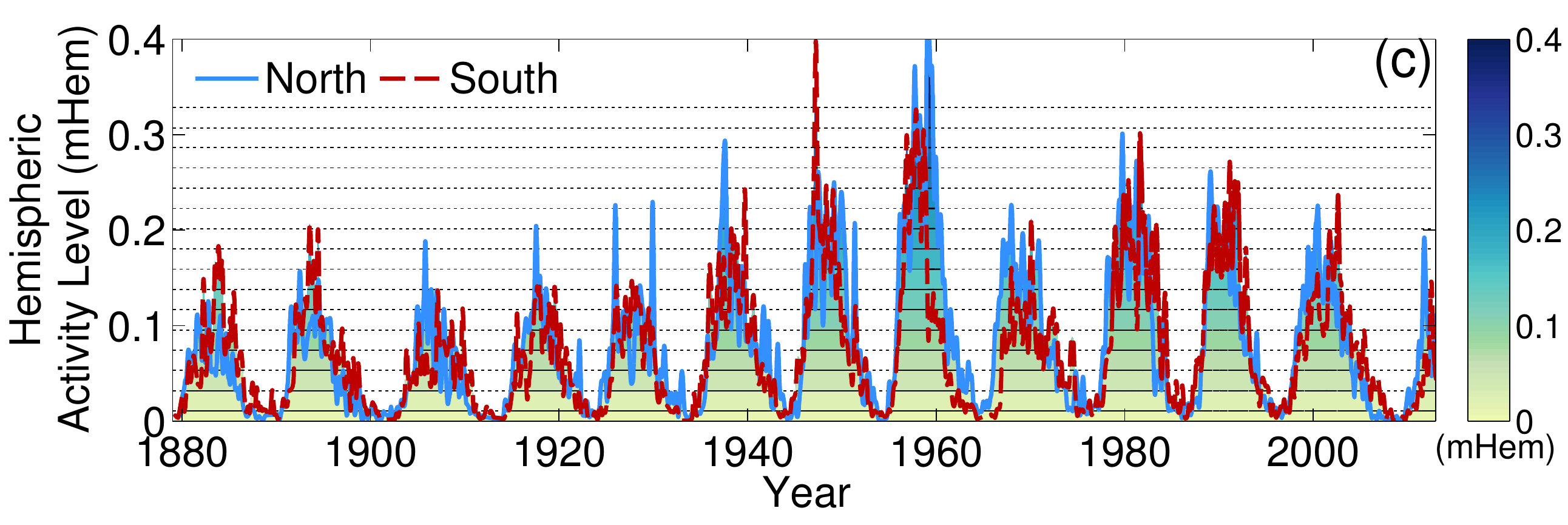}\\
  \includegraphics[width=0.75\textwidth]{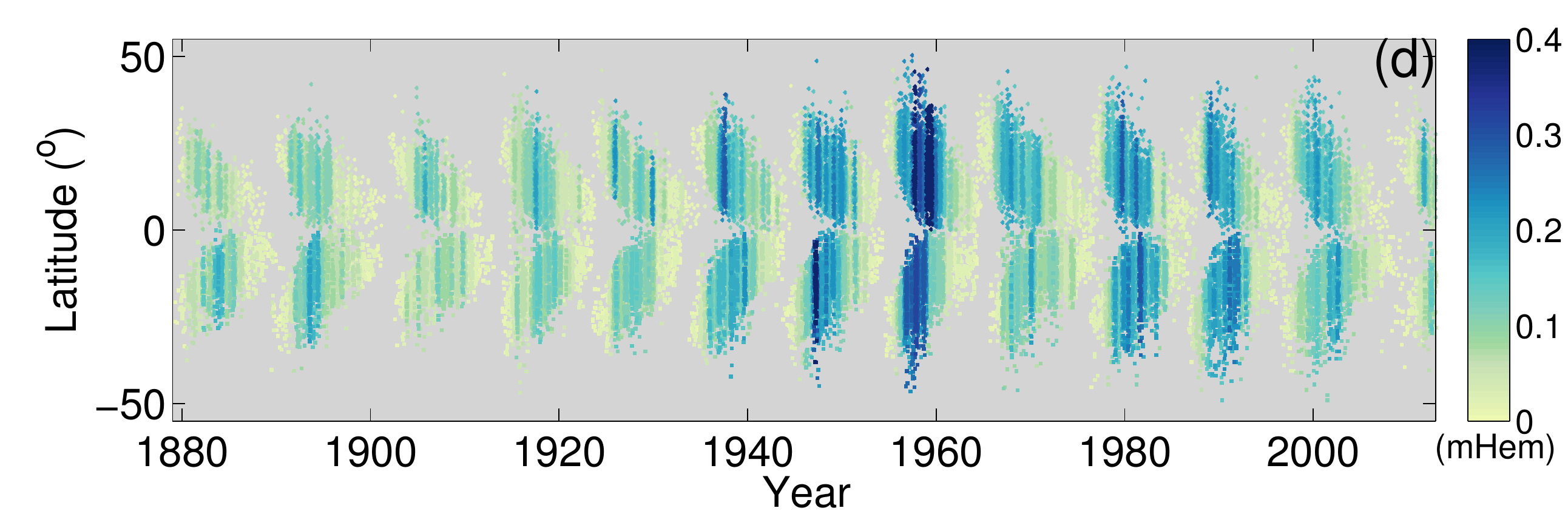}
\end{center}
\caption{(a)-(b) Temporal binning of sunspot data vs.\ (c)-(d) data binning by activity level.  In panels (a)-(c), northern (southern) hemispheric data are marked using a light blue (dark red) color.  In panels (c) and (d), activity level is indicated using a color scale changing from light yellow to dark blue.  Note that the vertical and color axes are the same in panel (c).  Activity level, displayed in panels (b) and (c), is calculated by convolving the daily sunspot group area with a six-month Gaussian filter.}\label{Fig_Bfly_Bin}
\end{figure*}

\section{Time: The Predominant Way of Binning Data}\label{Sec_Time_bin}

Given that the solar cycle is a continuously evolving transient process, time has been the traditional primary focus of statistical analyses of sunspot and BMR properties.  This means that comparative studies break down and compartmentalize data into chunks defined by when they occur.   This kind of binning is illustrated in Figures \ref{Fig_Bfly_Bin}(a) and (b), showing vertical markers that break down our data into separate cycles.  There are essentially three ways of binning data in time; to illustrate them, in this section we make a very limited review of studies using each type (which should not be considered comprehensive).  Example studies are chosen specifically to highlight the advantages of each approach.

\subsection{Binning Data by Arbitrary Time Interval}

The most straightforward way of time binning is to separate data into specific time intervals (by day, month, or year or by carrington rotation).  This kind of binning is the most natural form of data exploration \citep[see, for example, the pioneering studies by][]{tang-etal1984,wang-sheeley1989}, and it is still widely used today.   It is particularly powerful for searching for evidence of long-term trends (spanning timescales longer than the solar cycle itself).  More recent examples of this kind of binning are the thought-provoking papers by \cite{penn-livingston2006,penn-livingston2011} who reported a decrease in the average magnetic field of sunspots since 1998; they speculated that, if this trend continued, sunspots would become very rare by 2025.  Very good examples of how best to use this kind of binning are the responses to Penn \& Livingston by \cite{nagovitsyn-etal2012} and \cite{pevtsov-etal2014} where, using data going back to 1920, they demonstrate that there is no apparent long-term trend in the evolution of the average properties of sunspots; there is simply a cyclic modulation.

\subsection{Binning Data by Cycle}

The second way of binning data involves grouping data according to the cycle to which they belong.  This kind of binning is very good for identifying significant changes between different cycles.  Some examples of this kind of binning are the papers by \cite{hathaway-choudhary2008} who looked at the decay rate of sunspots, \cite{mcclintock-norton2013} who studied variations in sunspot group tilt angles and Joy's Law, and \cite{detoma-etal2013a}, who examined changes in sunspot area between cycles 22 and 23.

This type of time binning truly excels when used for studying the physical mechanisms responsible for sustaining and propagating the solar cycle. An excellent example of this kind of work was performed by \cite{dasiespuig-etal2010} who by looking at cyclic averages of tilt angles, found a correlation between weighted averages of sunspot group properties during a cycle and the amplitude of the following cycle \citep[not to be confused with the reported same-cycle negative correlation between the amplitude of a cycle and tilt averages; see][]{ivanov2012,mcclintock-norton2013,dasiespuig-etal2013}.  This result has been further expanded by \cite{munoz-etal2013a} who demonstrated that this connection exists because the average properties of BMRs determine the strength of the poloidal field at the end of the solar cycle.  Together they represent observational evidence in favor of the BL mechanism and our current understanding of the solar cycle.

\subsection{Binning Data by Cycle Phase}

The last type of time binning we will review here is binning according to cycle phase.  This involves either choosing specific phases of the solar cycle (rising, maximum, declining, and minimum phases), or binning the data relative to its position within a particular solar cycle.  Some examples of this kind of binning are the papers by \cite{mathew-etal2007} who studied the dependence of umbral and penumbral brightness on the solar cycle, \cite{zharkova-zharkov2008} who examined daily variations of tilt angles during the rising and declining phases of cycle 23; and \cite{watson-fletcher-marshall2011} who looked at maximum magnetic field and umbral/penumbral areas during the different phases of cycle 23.

This type of time binning is very powerful for characterizing the general properties of the solar cycle.  An excellent example of this kind of work was performed by \cite{jiang-etal2011a}, who performed a very detailed quantitative characterization of the relationship between cycle amplitude, cycle phase, and the properties of active latitudes (i.e., the shape, location, and width of the wings in the butterfly diagram).  Taking advantage of this characterization, they laid a solid foundation for the construction of synthetic data sets based solely on sunspot number, which can be used to drive surface flux transport simulations.

\section{Activity Level:  A New Way of Binning Data}\label{Sec_AL_bin}

Although there is undeniable value in using time to bin sunspot and BMR data (when the objective is to study and characterize the solar cycle), this kind of binning is sub-optimal for studying how the evolution of the solar cycle changes the properties of BMRs and their associated sunspots.  The problem is conceptual: a single BMR and its associated sunspot group is believed to be the photospheric manifestation of a buoyant flux tube.  These tubes take roughly a month to travel through the convection zone \citep{fan2009,weber-etal2011} and have a typical lifetime (after eruption) of about a month.  Considering that the solar cycle involves decadal timescales, this means that from the perspective of each BMR, the magnetic cycle is a quasi-static process (i.e., time-independent during a BMRs' life cycle).  It is commonly believed that the total number of BMRs and their combined flux is a direct indication of the strength of the underlying toroidal field.  Furthermore, this magnetic field strongly determines the resulting properties of the emerged BMRs \citep[see a review by][]{fan2009}.  \emph{We propose that the amplitude of the solar cycle at any specific moment in time (which we assume to be directly related to the characteristics of the underlying toroidal field), is the true quantity determining the statistical properties of sunspots and BMRs}, and that the time dependence of these properties is simply a direct consequence of the evolution of this toroidal magnetic field.

\begin{figure*}[ht!]
\begin{center}
\begin{tabular}{c}
  \includegraphics[width=0.75\textwidth]{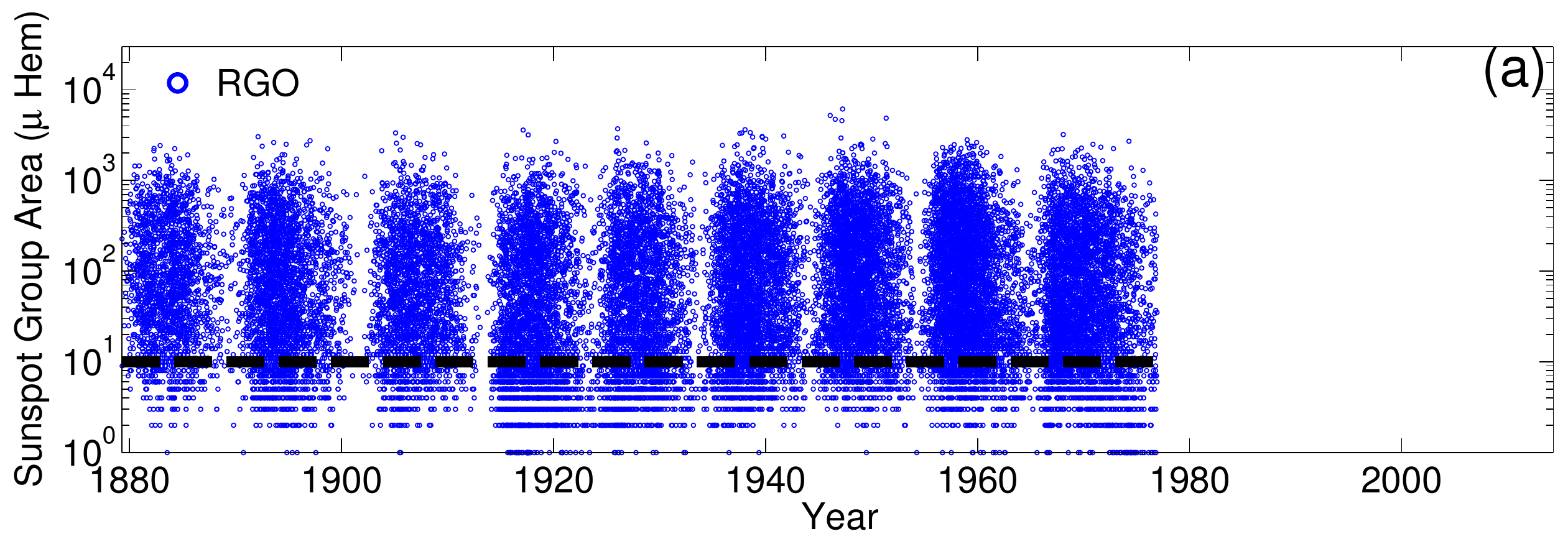}\\
  \includegraphics[width=0.75\textwidth]{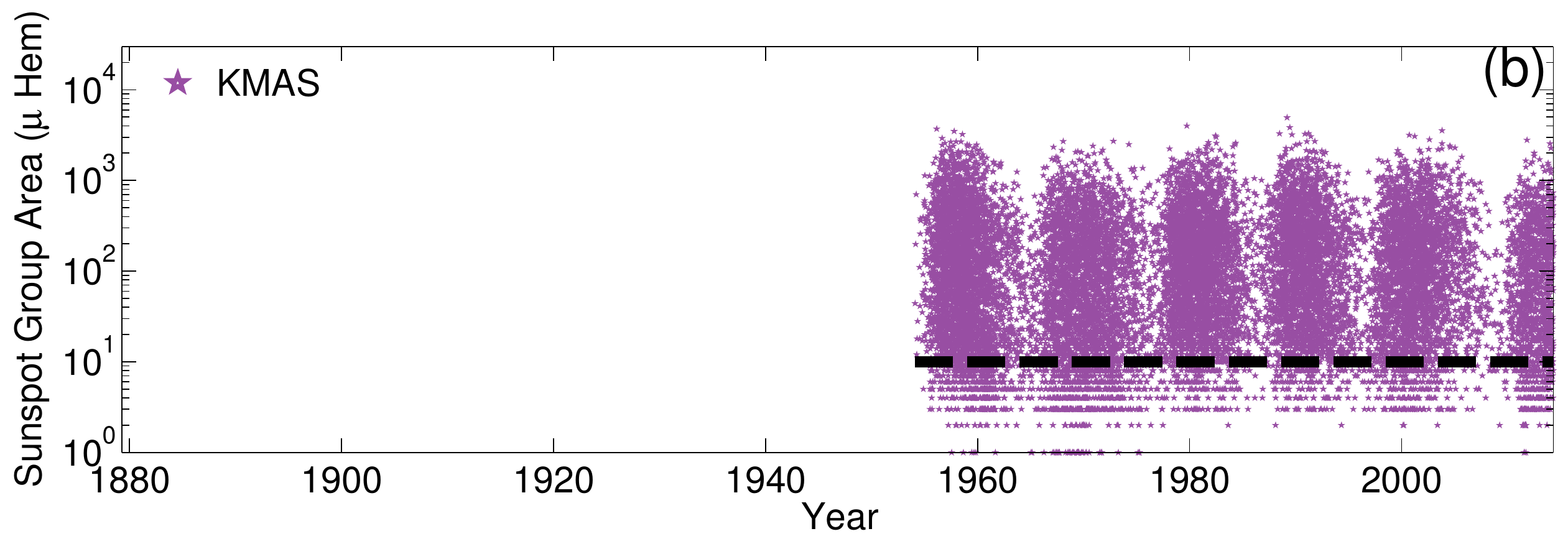}
\end{tabular}
\end{center}
\caption{Logarithmic plot of sunspot group size of the (a) RGO and (b) KMAS data sets.  Dashed black horizontal lines indicate the threshold above which data is used to calculate activity level and also fitted to the composite Weibull plus log-normal distribution.  This threshold is set an order of magnitude above the smallest structure of each set.}\label{Fig_Data}
\end{figure*}

To avoid possible misunderstandings, from now on we will use the term \emph{activity level} to refer to the mean level of hemispheric solar activity at any specific moment in time.  We do this in order to avoid confusion with a cycle's peak amplitude.  In order to bin our data according to activity level, we first calculate the total hemispheric daily sunspot area, remove high-frequency components by convolving it with a six-month gaussian filter \citep{hathaway2010,munoz-etal2013a}, and use the result to assign an activity level to each data point.  This kind of binning is illustrated in Figure \ref{Fig_Bfly_Bin}(c), showing horizontal lines demarcating different activity levels.  Note that activity level is calculated separately for each hemisphere.  It is also important to mention that we are assuming no intrinsic differences in sunspot and BMR properties across different hemispheres or across different cycles.  Instead, we assume that what makes each cycle unique is the actual evolution of activity level in each hemisphere (i.e., its ups and downs).

Figure \ref{Fig_Bfly_Bin}(c) and Figure \ref{Fig_Bfly_Bin}(d), showing a butterfly diagram in which each point is colored according to activity level, highlight some of the subtle but important aspects of binning by activity level.  First, we are assuming that activity levels and their associated statistical properties are not unique to any given cycle.  This means that sunspots and BMRs that appear during the absolute maximum of a weak hemispheric cycle have the same statistical properties as those that appear at the same activity level in stronger hemispheric cycles.  Cycle 19, the strongest cycle ever observed, is an extreme example of this (as it samples the entire range of activity levels we have observed so far).  Second, under this assumption, sunspots that appear simultaneously in the northern and southern hemispheres will have different statistical properties if there is a hemispheric asymmetry in activity level.\\

\begin{figure*}[ht!]
\begin{center}
\begin{tabular}{ccc}
  \includegraphics[width=0.3\textwidth]{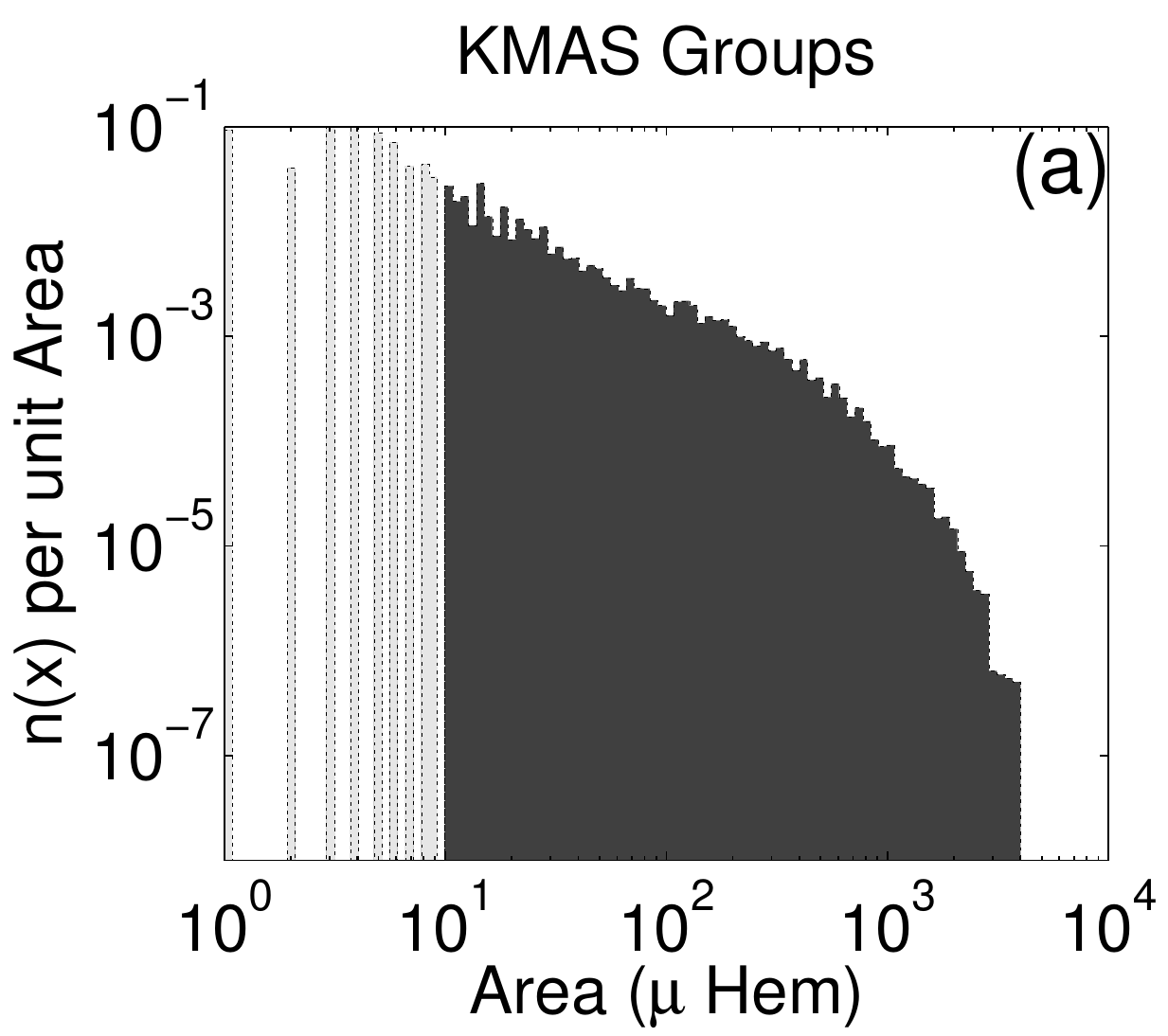} & \includegraphics[width=0.3\textwidth]{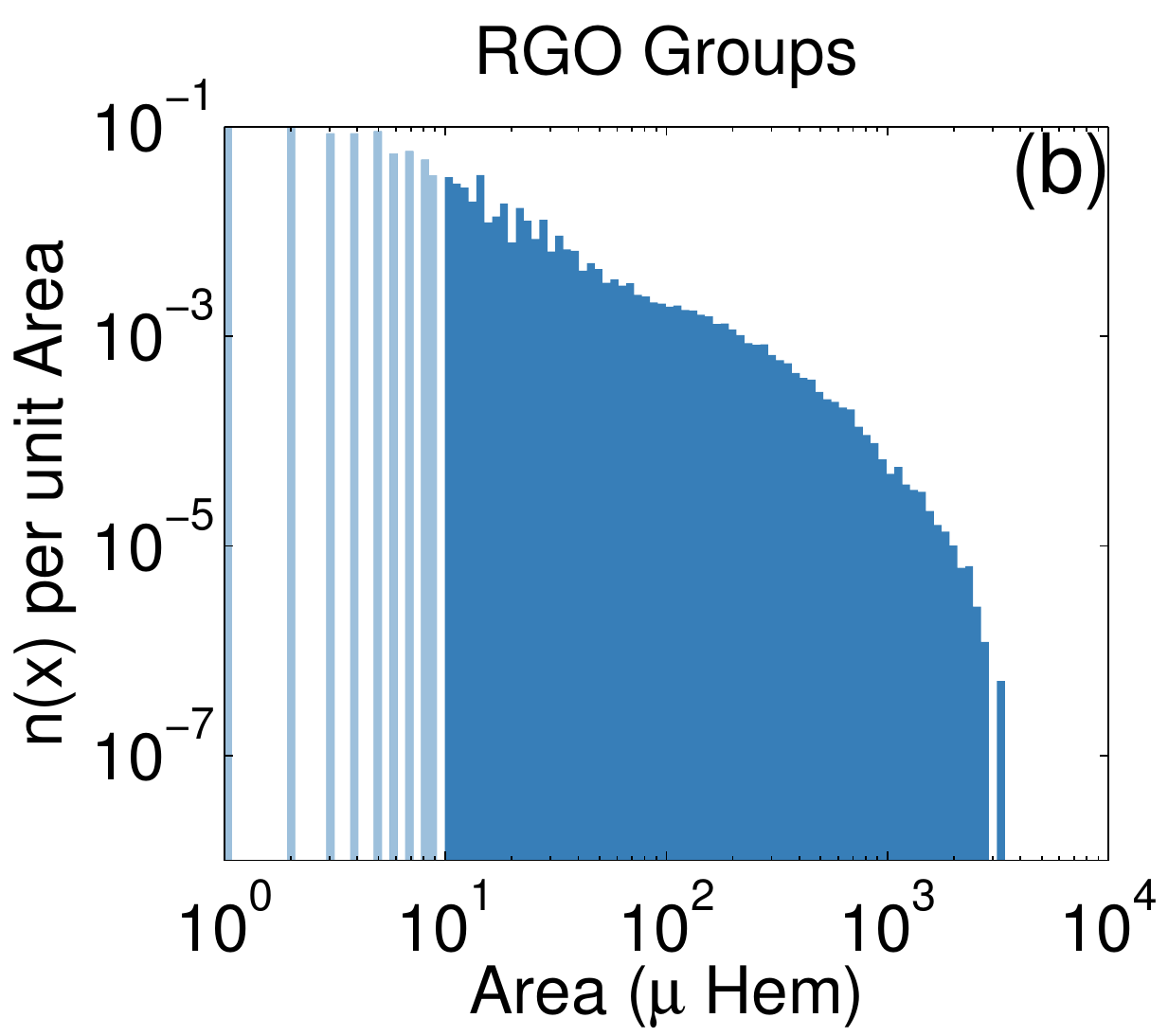} & \includegraphics[width=0.3\textwidth]{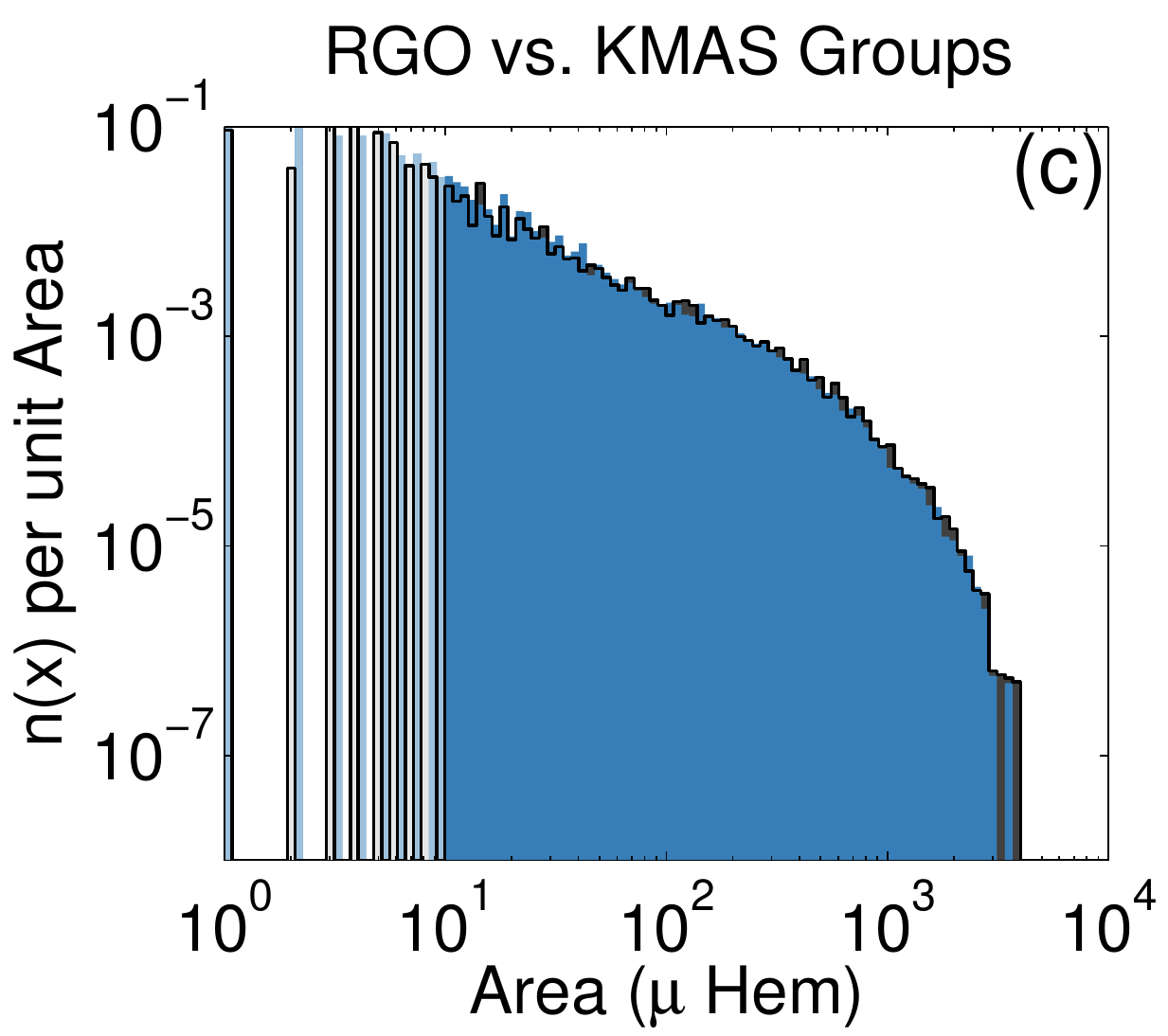}
\end{tabular}
\end{center}
\caption{(a) KMAS and (b) RGO empirical distribution functions for the period between 1954 and 1976.  (c) The multiplication of RGO sunspot group area by a factor of 1.06 maximizes the agreement between both empirical distribution functions.  Empirical distribution functions show all data in each set, but only data in the dark shaded region are used in the cross-calibration.}\label{Fig_Data_Fit}
\end{figure*}

\section{Data Selection and Truncation}\label{Sec_Data}

As the backbone of our analysis we use the sunspot group database compiled and published by the Royal Greenwich Observatory (RGO).  This set includes heliographic positions and areas of sunspot groups observed from 1874 to 1976 by a small network of observatories: the Cape of Good Hope, Kodaikanal and Mauritius. The RGO data, covering nine solar cycles (from cycle 12 to cycle 20), provide the longest and most complete record of sunspot group areas. We extract from this database a single area and position for each sunspot group.  We assign to the group the single largest reported area in all its days of observation.  The result is a set of 30,026 groups. Data are shown in Figure \ref{Fig_Data}(a).

Since part of our objective is to study the transition between sunspot cycles 23 and 24, we supplement the RGO data using observations taken by the Kislovodsk Mountain Astronomical Station (KMAS) of the Central Astronomical Observatory at Pulkovo.  The KMAS has been in continuous operation since 1948, making it one of the very few institutions performing a wide array of solar surveys through the entirety of the space age.  This makes it quite valuable as a connecting set between modern missions and previous surveys.  This database contains 108,364 sunspot group observations taken from 1954 February 9 to the present (covering 6.5 solar cycles, from cycle 18 to cycle 24), giving us a nice overlap with the RGO set that we can use to cross calibrate-them.  As with the RGO set, we extract a single area and position for each sunspot group.  We assign to the group the single largest reported area in all its days of observation.  The result is a set of 19,221 groups.  KMAS data are available at \href{http://158.250.29.123:8000/web/Soln_Dann/}{http://158.250.29.123:8000/web/Soln\_Dann/}. Data are shown in Figure \ref{Fig_Data}(b).

As recounted in detail by \cite{munoz-etal2015a}, there is a host of issues that can potentially distort the statistical properties of structures near the lower detection threshold.  They include but are not limited to observational bias, artificial binning caused by resolution, convolution of instrumental cadence and feature lifetime, underestimation due to the quality of the observing conditions, and underestimation due to excessive complexity in the observed phenomenon.   Considering that small structures are also the most numerous, here we follow the suggestion by C.\ DeForest (2014, private communication) and implemented by \cite{munoz-etal2015a}, of imposing a truncation limit one order of magnitude above the minimum size of detection in our databases. We only use data above this limit in our distribution fits and analysis.  The location of these thresholds, shown in Figure \ref{Fig_Data} as dark horizontal lines, successfully isolates problematic data from the rest of each set.

\subsection{Cross-Calibration}

\begin{figure}[ht!]
\begin{center}
  \includegraphics[width=0.35\textwidth]{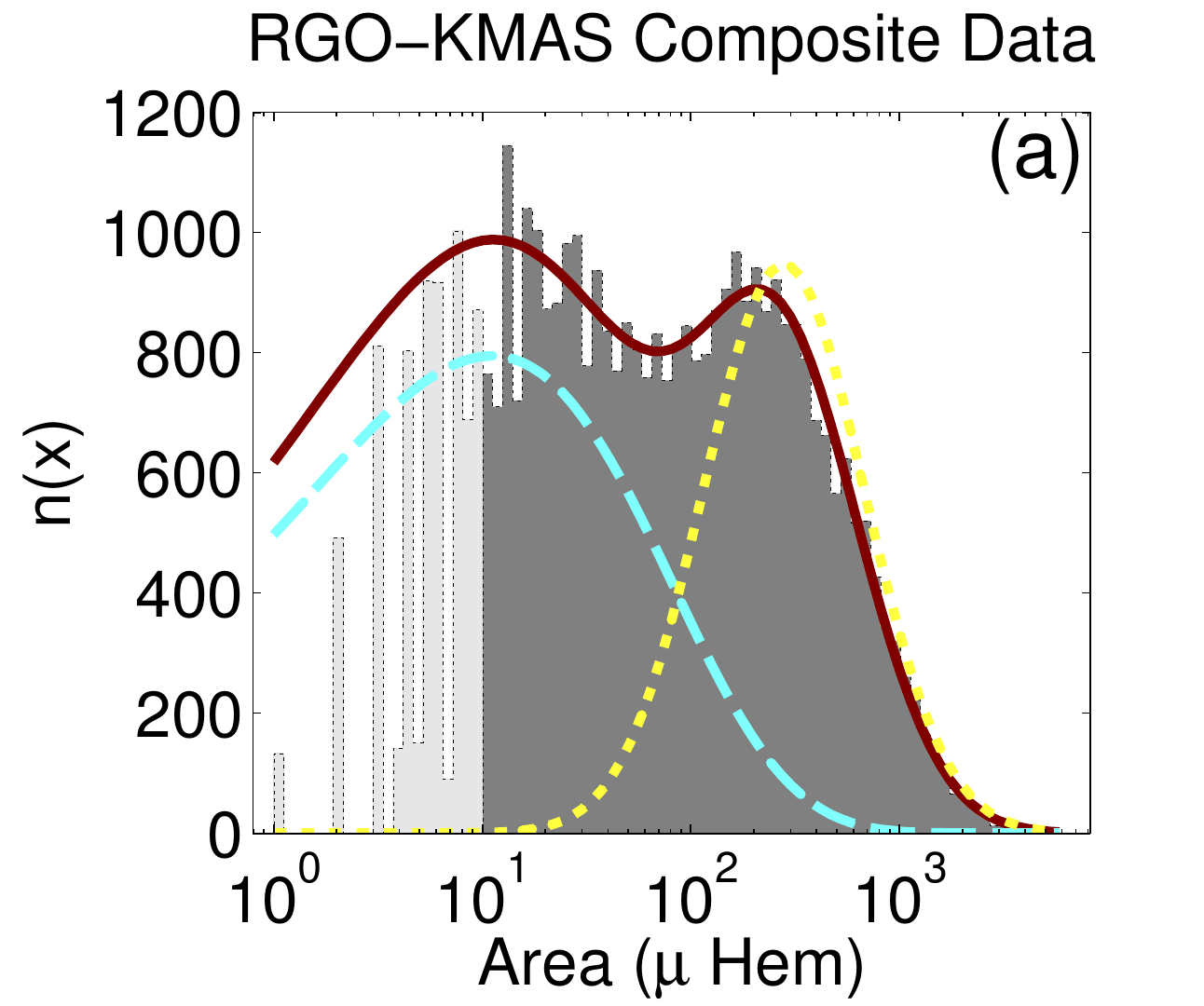}\\
  \includegraphics[width=0.35\textwidth]{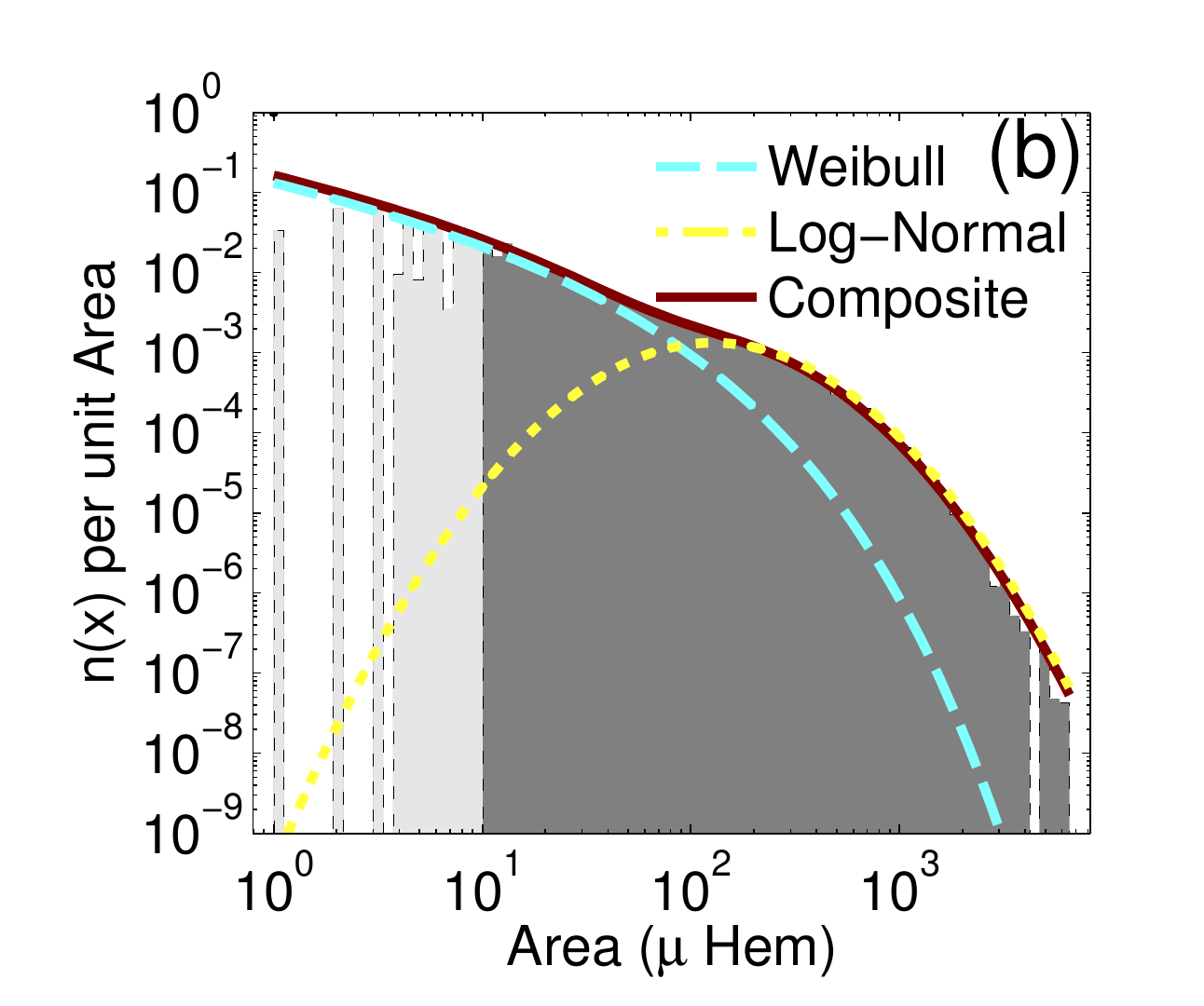}
\end{center}
\caption{(a) Histogram using logarithmic binning of RGO/KMAS data. (b) Empirical PDF of RGO/KMAS data.  Both represent different ways of looking at the same data, and both panels are overplotted with the same fit using a linear combination of Weibull (dashed blue line) and log-normal distributions (dotted yellow line). The composite fit is shown as a solid dark red line.   Both panels include all data in the set, but only data shown in a dark shade are included in the fit.}\label{Fig_Emp_Dist}
\end{figure}

Our first task is to cross-calibrate the RGO and KMAS data sets to form a composite spanning from 1874 to the present.  Given that the KMAS survey is still active, we use it as our reference set.  This way we will be able to extend our composite database into the future as KMAS continues to perform observations.  Since for this study it is important for the sets to be statistically compatible, we find the proportionality calibration constant by matching the KMAS (Fig.~\ref{Fig_Emp_Dist}(a)) and RGO (Fig.~\ref{Fig_Emp_Dist}(b)) empirical size distribution functions.  For this purpose, we use only data belonging to the overlapping interval between the RGO and KMAS data sets (i.e. between 1954 and 1976).  This technique was used by \cite{munoz-etal2015a} to reconcile and cross-calibrate 11 different sunspot group, sunspot, and bipolar magnetic flux data sets. It involves the following steps:

\begin{enumerate}
  \item Choose a proportionality constant out of a range of possible values.
  \item Multiply all sunspot group areas in the RGO database by this proportionality constant (effectively shifting the empirical distribution left or right in logarithmic scale).
  \item Evaluate if the resulting empirical distribution overlaps with the reference KMAS empirical distribution.
  \item Find the root mean square error (RMSE) between the overlaps.
  \item After trying all possible proportionality values in a set, identify which one minimizes the RMSE.
\end{enumerate}

We find that multiplying RGO data by a factor of $1.06$ maximizes the agreement between the KMAS and RGO empirical distribution functions (shown in Fig.~\ref{Fig_Emp_Dist}(c)). We construct our composite by using all RGO data (with areas multiplied by the $1.06$ factor), and KMAS data from 1977 onward.

\section{Size Distribution, Distribution Fitting, and Model Selection}\label{Sec_Math}

In the past, different characterizations of the size-flux distribution of magnetic structures have used different statistical distributions to fit the data: the exponential distribution \citep{tang-etal1984,schrijver-etal1997}, the log-normal distribution \citep{bogdan-etal1988,baumann-solanki2005,zhang-wang-liu2010,schad-penn2010}, exponential polynomials \citep{harvey-zwaan1993}, the Weibull distribution \citep{parnell2002}, the power-law distribution \citep{meunier2003,zharkov-zharkova-ipson2005,parnell-etal2009}, and linear combinations of these distributions \citep{kuklin1980,jiang-etal2011a,nagovitsyn-etal2012,munoz-etal2015a}.

In this work we build upon the results of \cite{munoz-etal2015a} who found after an in-depth quantitative comparison between the different proposed distributions that a linear combination of Weibull and log-normal distributions fits observations best.  This linear combination is used to define the probability-density function (PDF) of sunspot group sizes:
\begin{align}\label{Eq_Mix}
    f(x;k,\lambda,\mu,\sigma, c) = {} &  \frac{(1-c)k}{\lambda}\left(\frac{x}{\lambda}\right)^{k-1} e^{-(x/\lambda)^k}\nonumber \\
                                      & + \frac{c}{x\sigma\sqrt{2\pi}}e^{ -\frac{(\ln x-\mu)^2}{2 \sigma^2} },
\end{align}
where $x$ is the area of a given magnetic structure, $k>0$ and $\lambda>0$ are the shape and scale parameters of the Weibull distribution, $\mu$ and $\sigma$ are the logarithmic mean and standard deviation characterizing the log-normal, and $0\leq c\leq1$ is the proportionality constant that blends these distributions together.  Note that we introduced a small change with respect to the composite distribution defined by \cite{munoz-etal2015a}: here, the Weibull (log-normal) term is multiplied by $1-c$ ($c$). It is also important to highlight that Equation (\ref{Eq_Mix}) is normalized so that its integral over the entire space is equal to one.  This is necessary so that we can later compare the empirical and analytical PDFs associated with activity level bins that contain different amounts of data points.

Due to the fact that we are working with truncated sets, we use the following truncated form of our PDF on our fits:
\begin{equation}\label{Eq_PDFtrunc}
    \operatorname{PDF}_{\text{trunc}}(x) = \frac{\operatorname{PDF}(x)}{1-\operatorname{CDF}(x_{\text{trunc}})},
\end{equation}
where CDF denotes the cumulative distribution function associated with Equation (\ref{Eq_Mix}), and $x_{\text{trunc}}$ denotes the limit value below which data is not used in the fit (see Section \ref{Sec_Data}).

In order to fit this PDF to the data we use maximum likelihood estimation (MLE).  Its basic idea is to find the set of parameters that maximizes the likelihood of a statistical model given the observed data.  For this purpose, the user defines and maximises a likelihood function constructed using the probability of observing all data in the set.  This method is far superior to fitting functional forms to histograms because it is not sensitive to the details of data binning.  A more detailed description of MLE can be found in Appendix \ref{Sec_MLE}, and in most modern statistics books \citep[for example in][]{hoel1984}.

\begin{table}[ht!]
\begin{center}
\begin{tabular*}{0.45\textwidth}{@{\extracolsep{\fill}}  c c c c c}
\multicolumn{5}{c}{\textbf{Composite Fit to RGO/KMAS data}}\\
\toprule
\multicolumn{2}{c}{Weibull} & \multicolumn{2}{c}{Log-Normal} &         \\
      k    &  $\lambda^*$   & $\mu$       & $\sigma$         & c       \\
  0.48     &  11.14         & 5.63        & 0.88             & 0.55\\
$\pm$0.15  &  $\pm$3.98     & $\pm$0.13   & $\pm$0.04        & $\pm$0.10
\end{tabular*}
\end{center}
\hspace{1em}
  \caption{Fitting parameters of the composite distribution to RGO/KMAS sunspot group data.  Quantities accompanied by a $^*$ are in units of $\mu$Hem, and other quantities are dimensionless.  The first row contains the fitted parameters; the second row the values of their 95\% confidence intervals.}\label{Tab_Mix}
\end{table}

To quantify the relative improvement of our statistical model by separating our data according to activity level, we use Akaike's information criterion \citep[AIC;][]{akaike1983}.  The AIC is a powerful tool for discriminating between different models by making an estimate of the expected, relative distance between the fitted model and the unknown true mechanism that generated the observed data. It uses a combination of the likelihood of the data and the model's degrees of freedom (dof) to strike a balance between bias and variance (i.e., between underfitting and overfitting).  A more detailed description of AIC can be found in Appendix \ref{Sec_AIC} and in an excellent book by \cite{burnham-anderson2002}.

\begin{figure*}[ht!]
\begin{center}
\begin{tabular}{ccc}
  \includegraphics[width=0.3\textwidth]{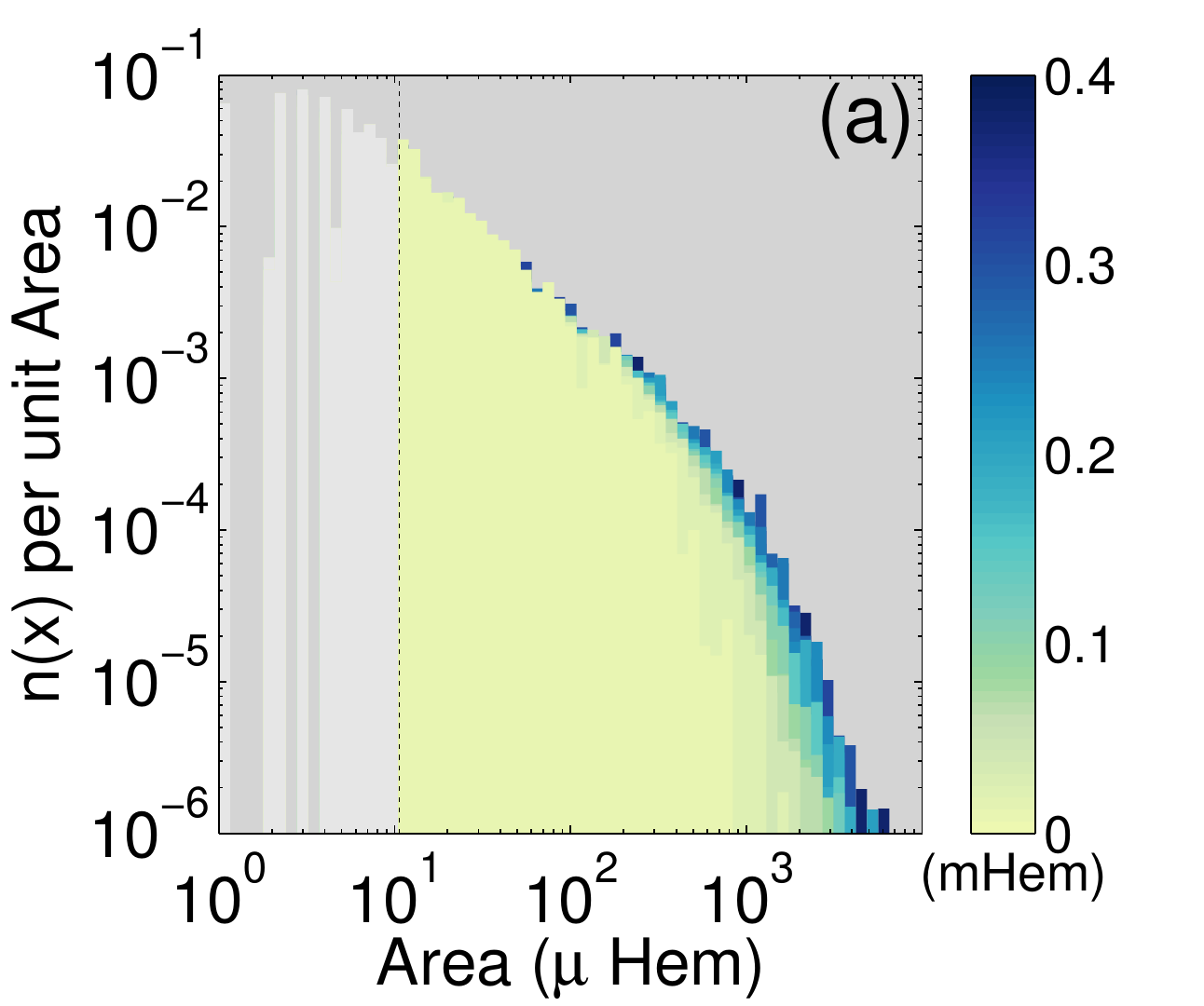}&\includegraphics[width=0.3\textwidth]{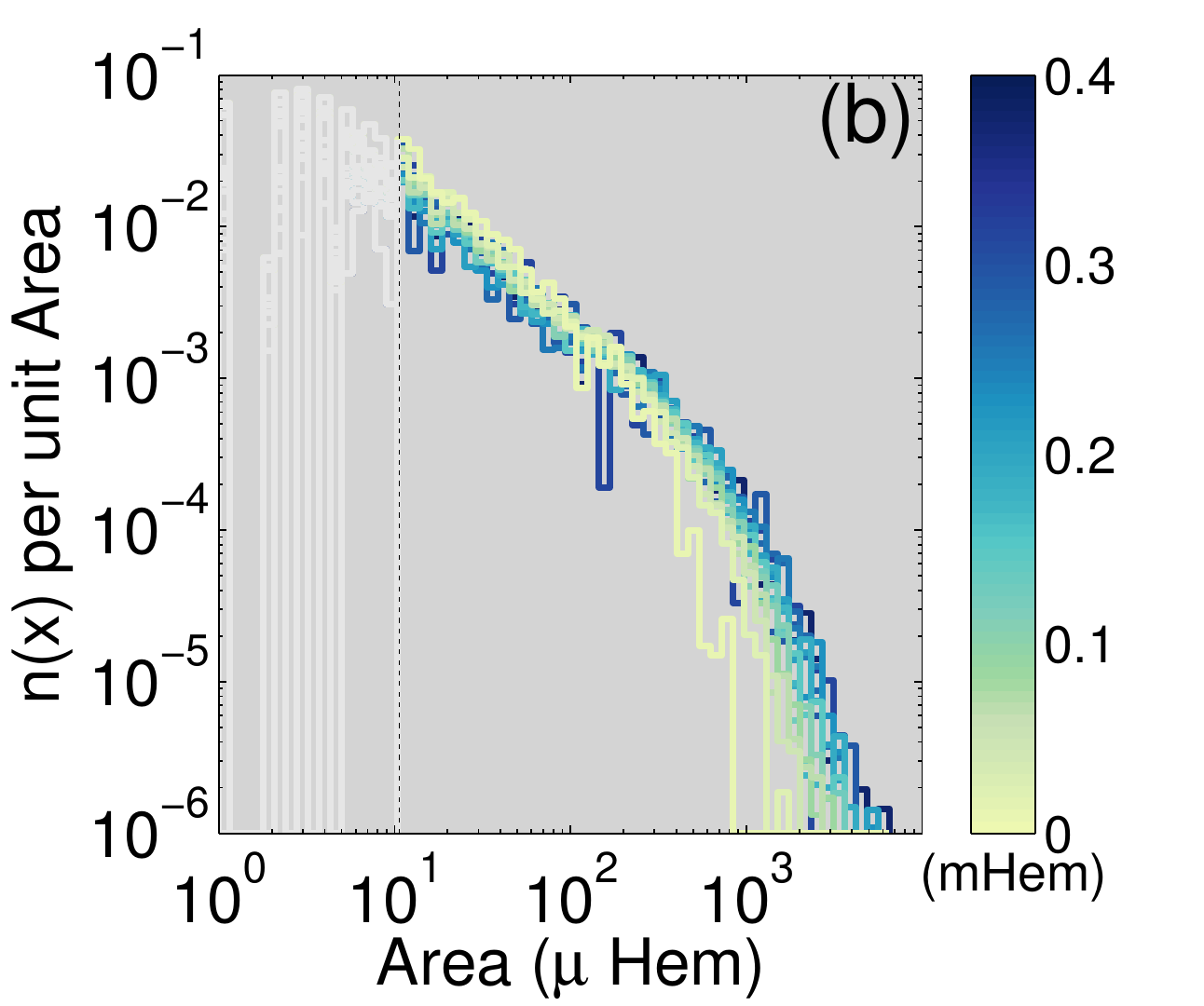}&\includegraphics[width=0.3\textwidth]{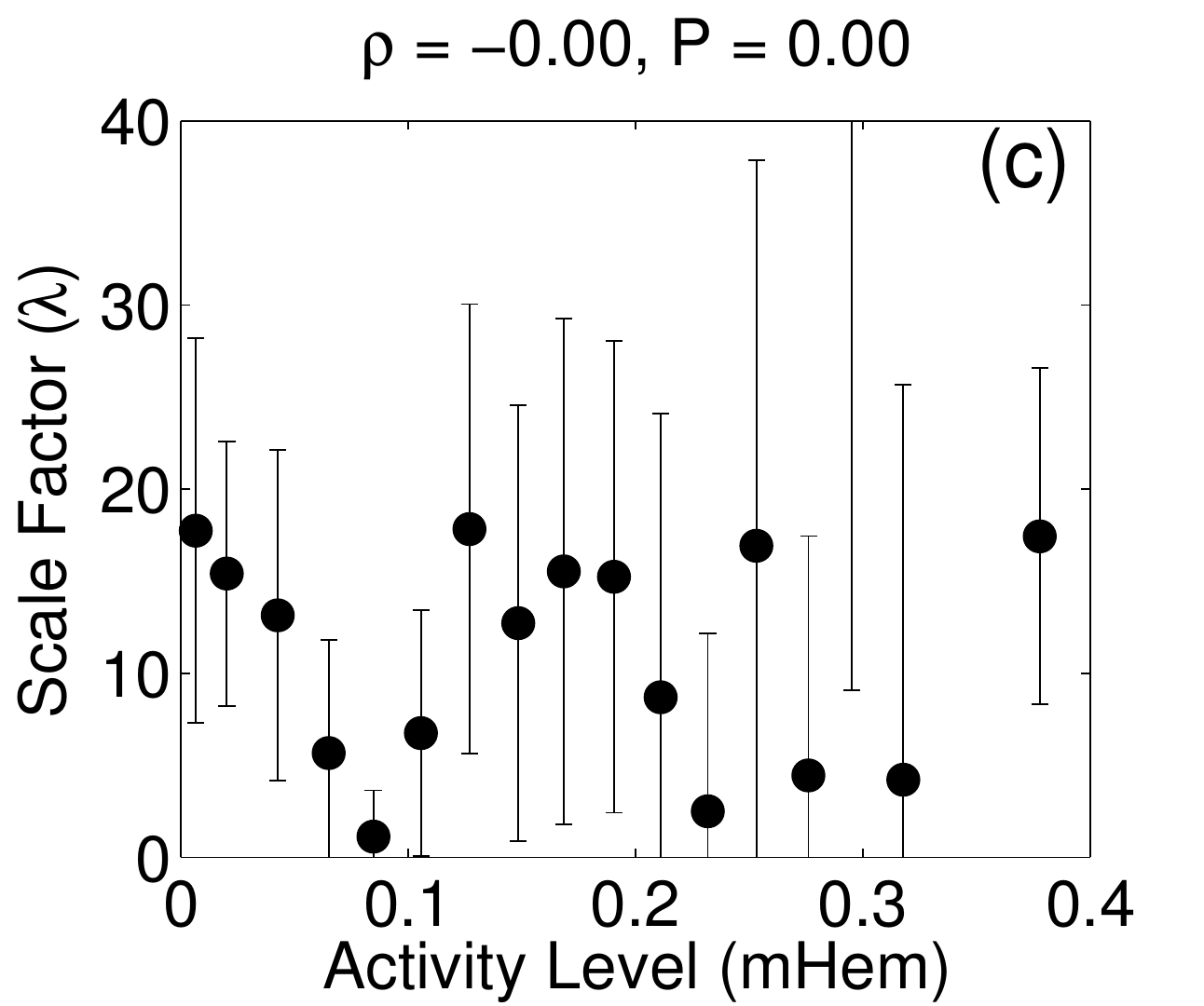}\\
  \includegraphics[width=0.3\textwidth]{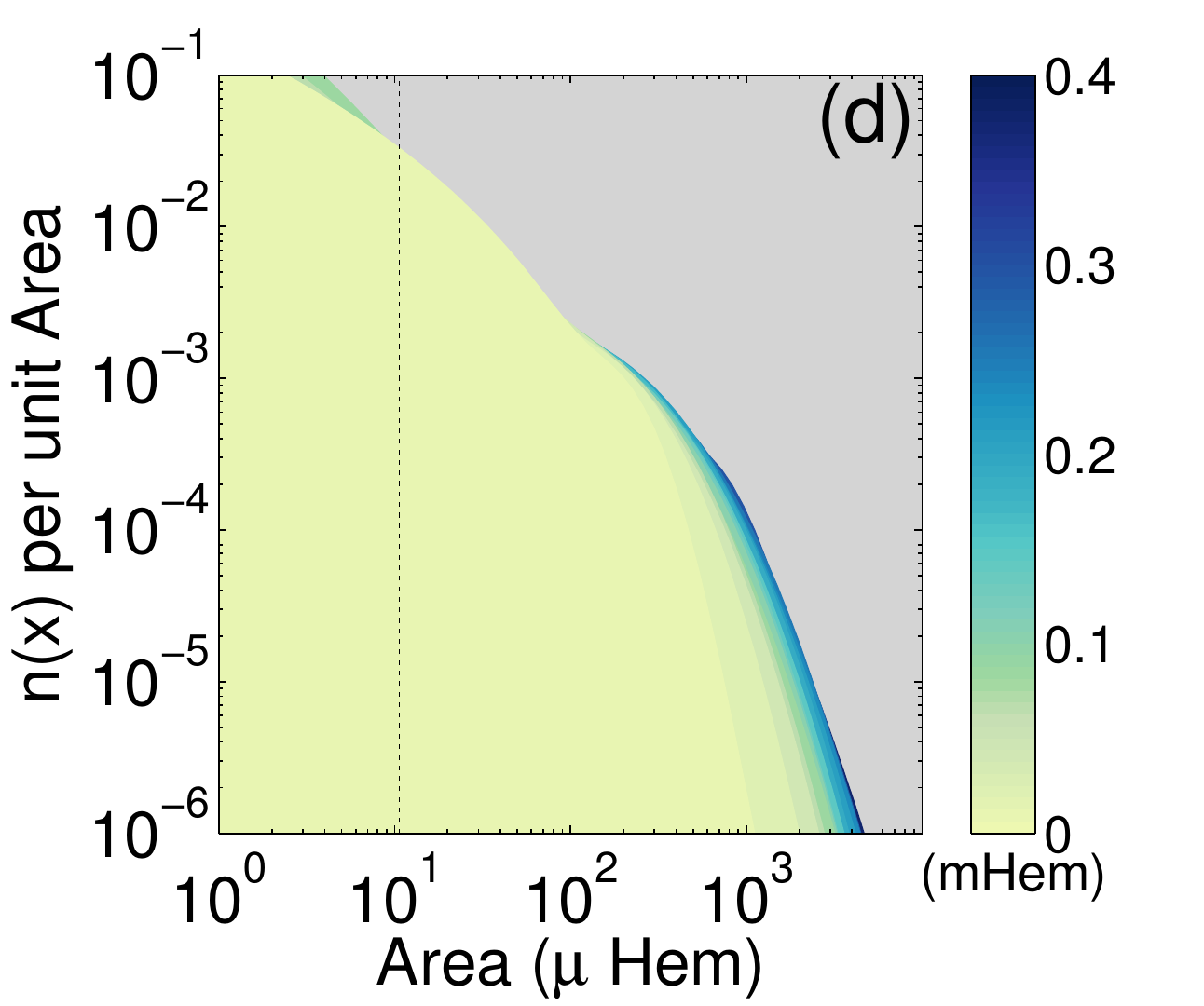}&\includegraphics[width=0.3\textwidth]{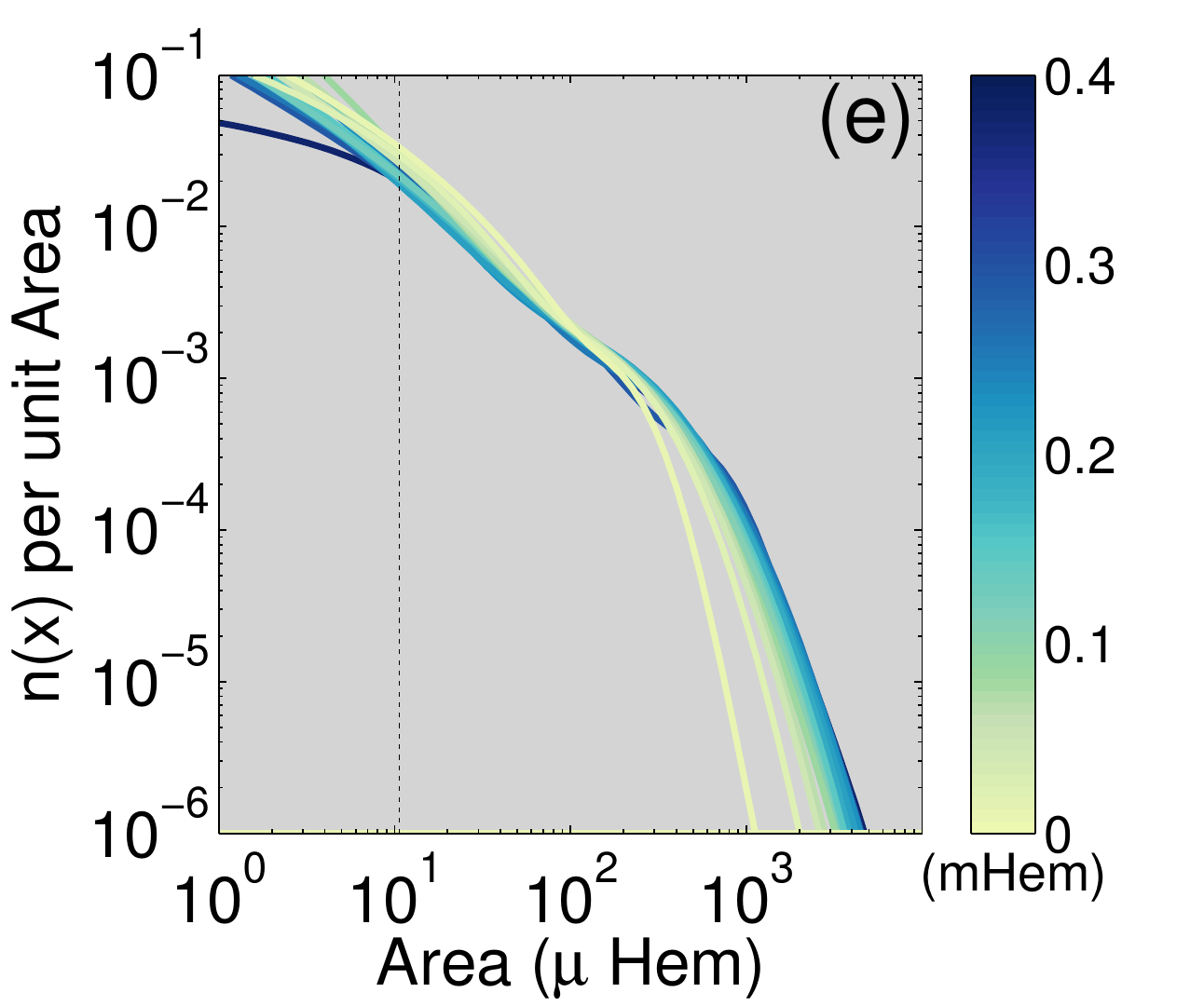}&\includegraphics[width=0.3\textwidth]{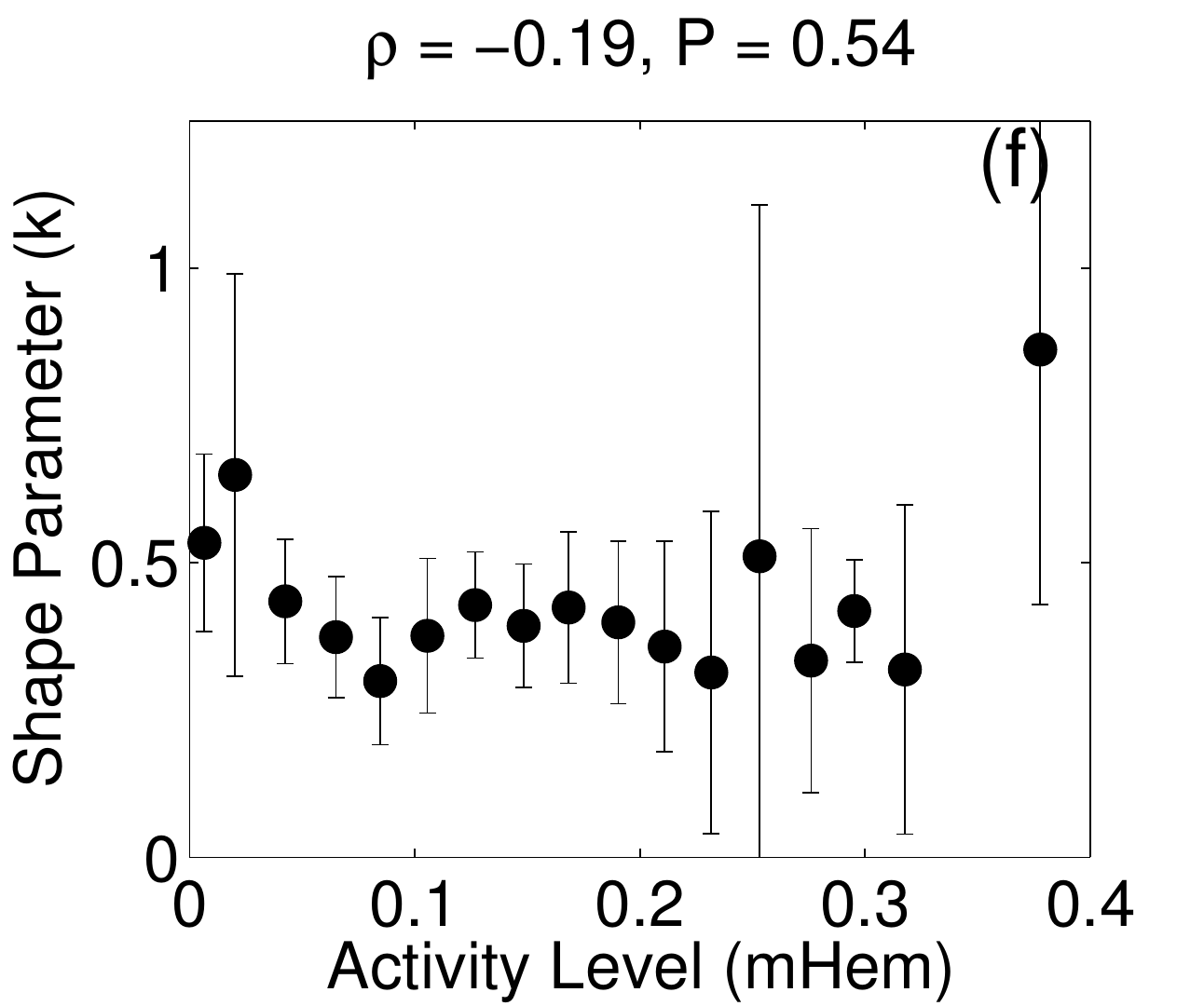}\\
  \includegraphics[width=0.3\textwidth]{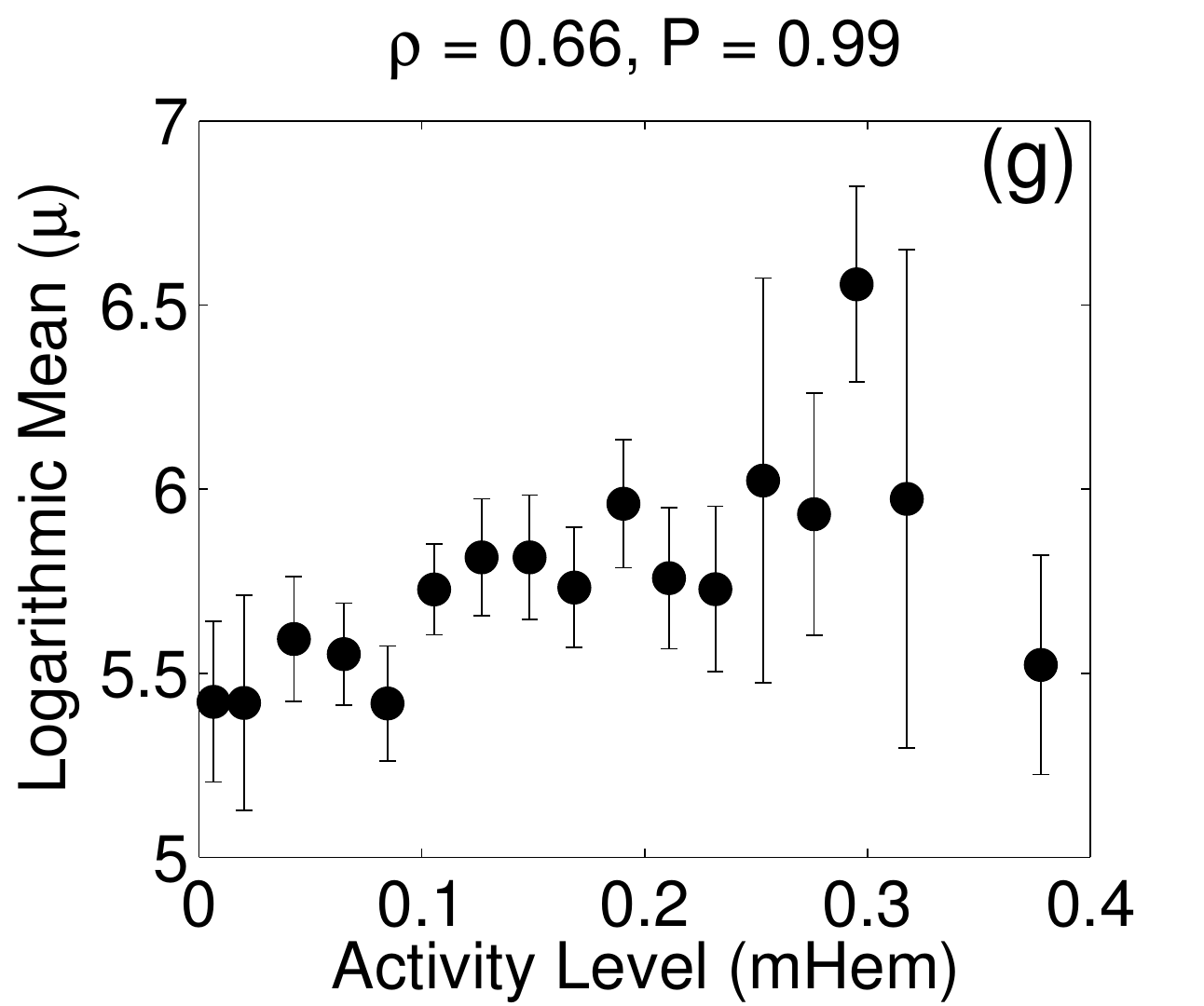}&\includegraphics[width=0.3\textwidth]{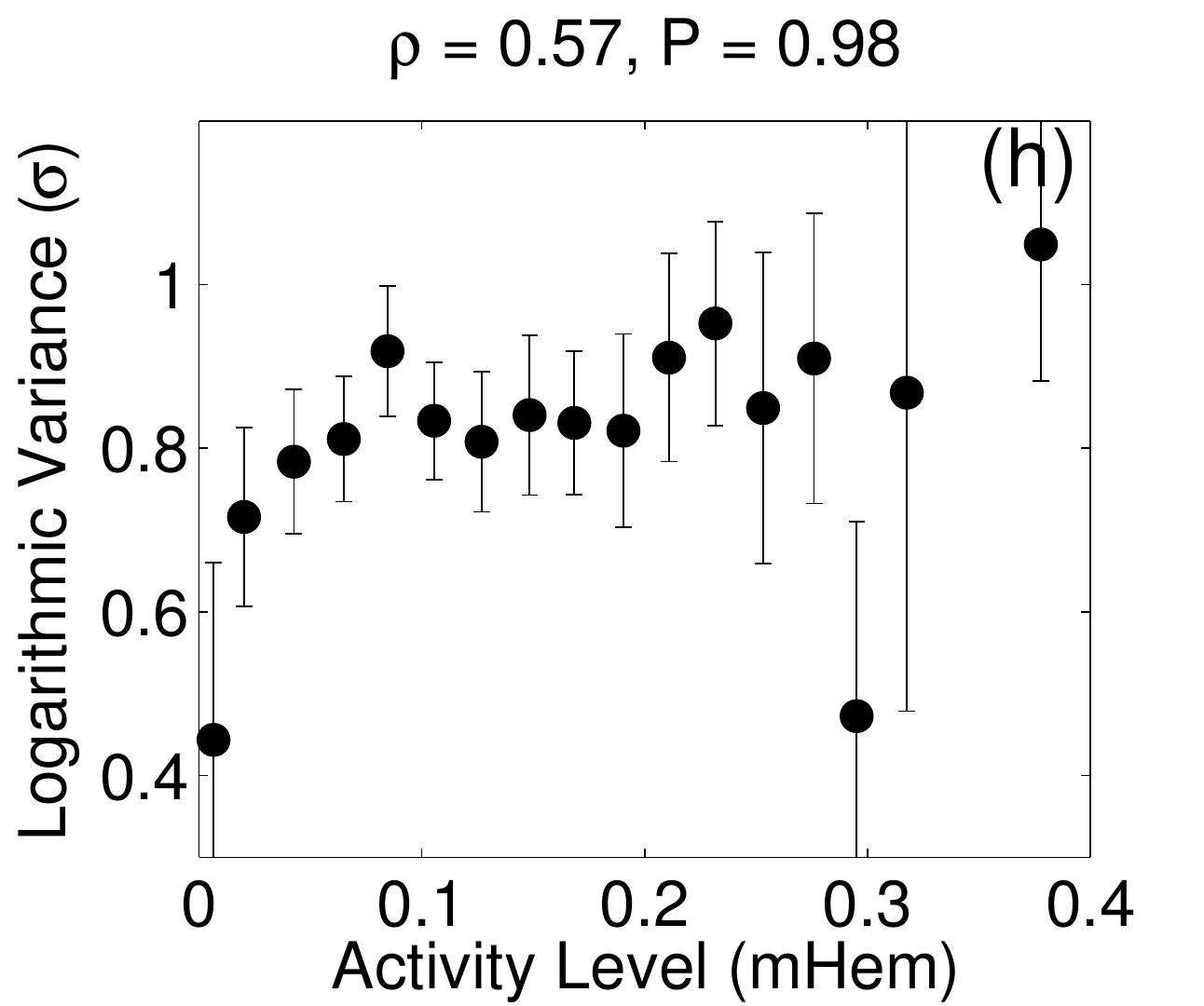}&\includegraphics[width=0.3\textwidth]{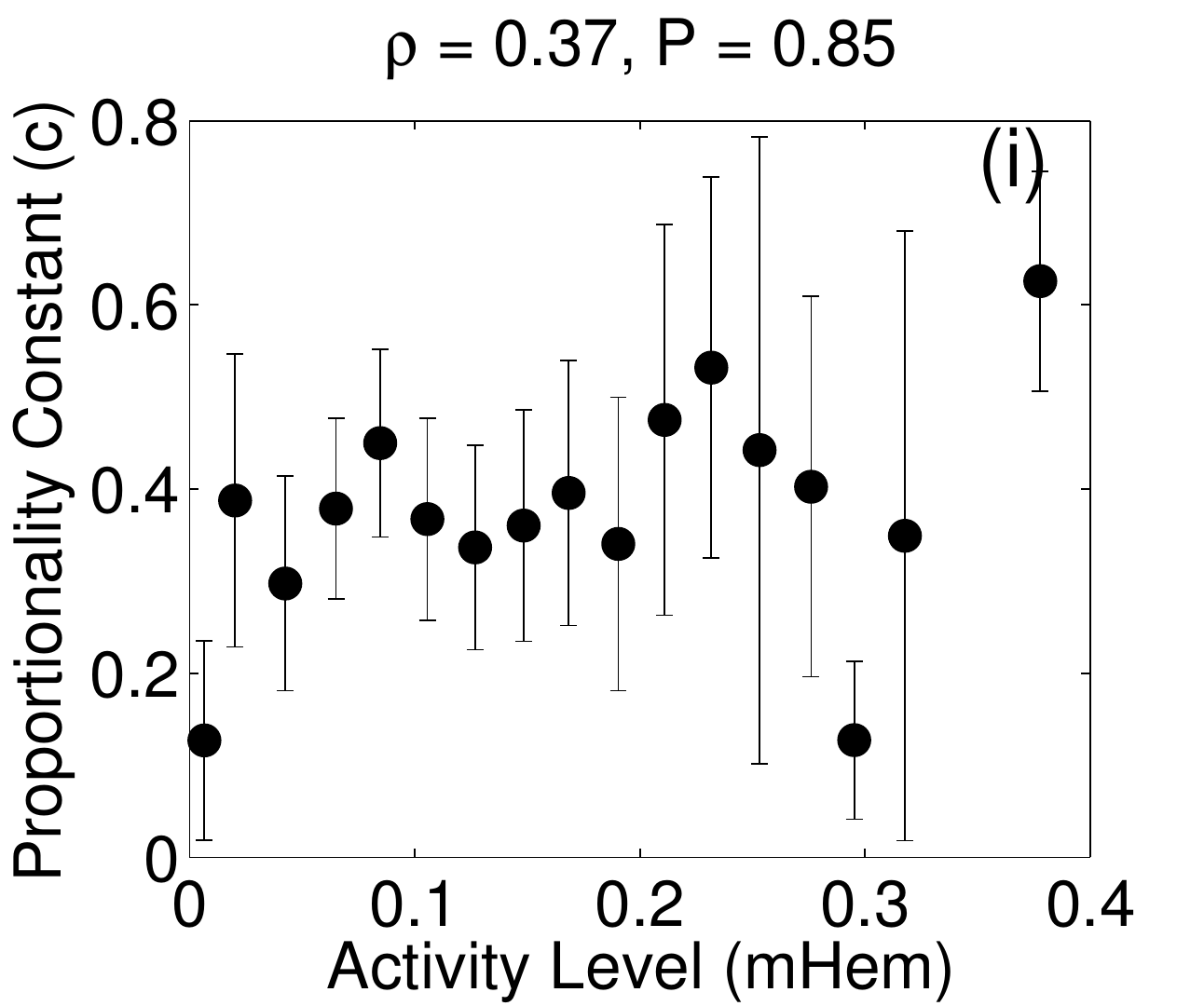}\\
\end{tabular}
\end{center}
\caption{(a), (b) Empirical size-distribution associated with different activity levels. (d), (e) Composite PDFs fitted to data binned according to activity level. In order to enhance perception, we use two ways of displaying the PDFs.  In (a) and (d), a solid color fills the area below each PDF and those associated with lower activity levels are plotted closer to the foreground.  In (b) and (e), each PDF is plotted using a thin line colored according to activity level.  Vertical dashed lines mark the limit below which data are not included in the fits.  Panels (c), (f)-(i) show the relationship between the different fitting parameters in the composite PDF and activity level (see Equation (\ref{Eq_Mix})).  Error bars indicate the 95\% confidence intervals of each value.  The Spearman's rank correlation coefficient ($\rho$) and its confidence level ($P$) are included as the title of each of these panels.}\label{Fig_Fits1}
\end{figure*}

\subsection{Fitting the Composite Distribution to the Entire RGO/KMAS Set}\label{Sec_Same_Dis}

Our first task is to fit the composite PDF to our RGO/KMAS database.  This helps us place this work in the light of the results of \cite{munoz-etal2015a} and gives us a reference against which we can evaluate the performance of PDF fitting of data binned by activity level.   The results of the fit are shown in Figure \ref{Fig_Emp_Dist}, and tabulated in Table \ref{Tab_Mix}.  They are in agreement with those found by \cite{munoz-etal2015a}.

\section{Fitting the Composite Distribution to the Binned RGO/KMAS Set}\label{Sec_Dis_Free}

After fitting our PDF to the entire data set, we now separate data according to activity level.  An inspection of the empirical size PDF functions associated with different activity levels (See Figures \ref{Fig_Fits1}(a) and (b)) shows a striking relationship between the abundance of large sunspot groups and activity level -- with groups bigger than 1000 $\mu$Hem being 30 times more likely to occur during high activity levels (e.g., the peak of cycle 19) than during a typical solar minimum.  Additionally, as can be observed in Figures \ref{Fig_Fits1}(d) and (e), the Weibull-log-normal composite is able to capture the overall shape of the empirical PDF in every case.

\begin{figure*}[ht!]
\begin{center}
\begin{tabular}{ccc}
  \includegraphics[width=0.3\textwidth]{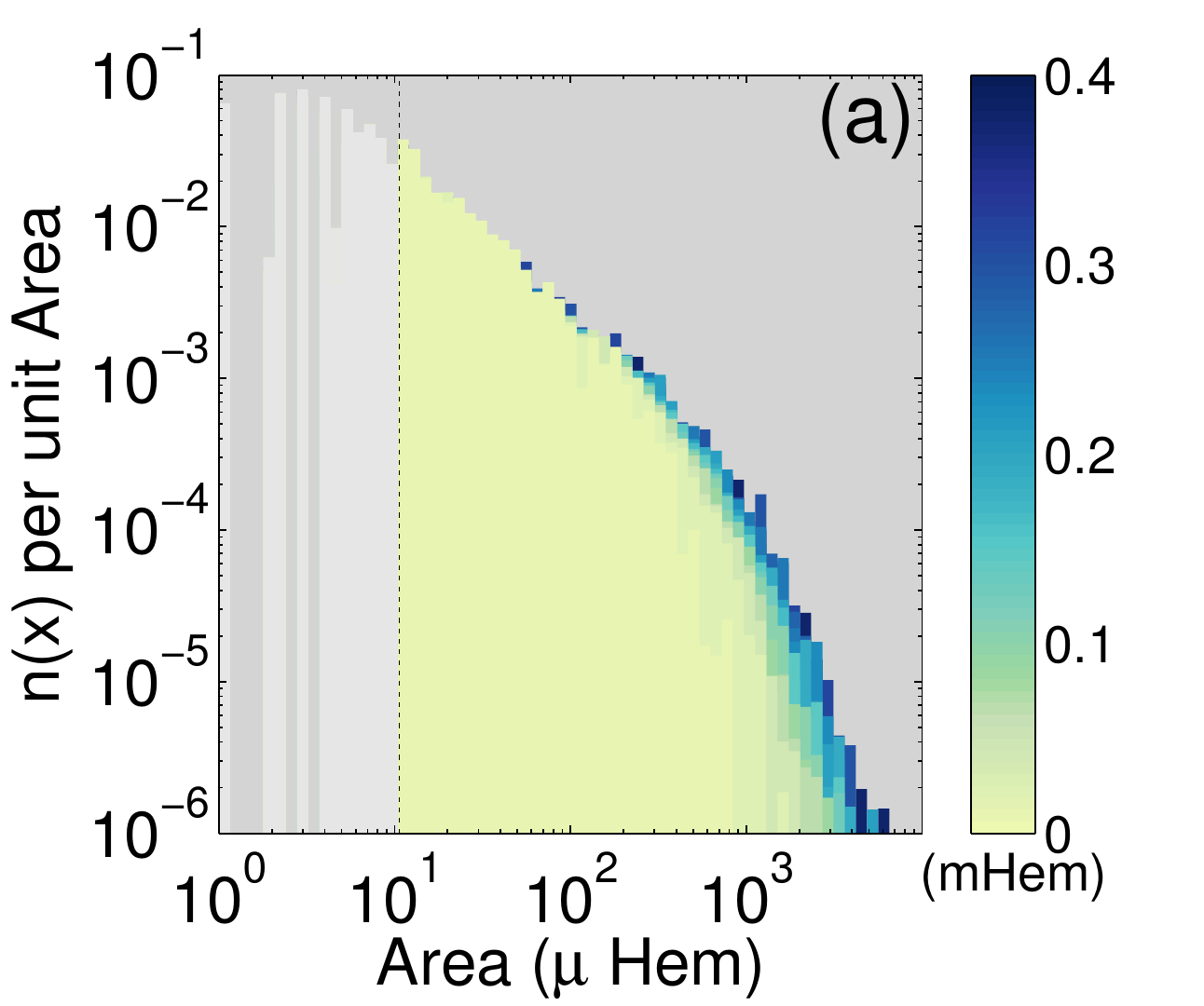}&\includegraphics[width=0.3\textwidth]{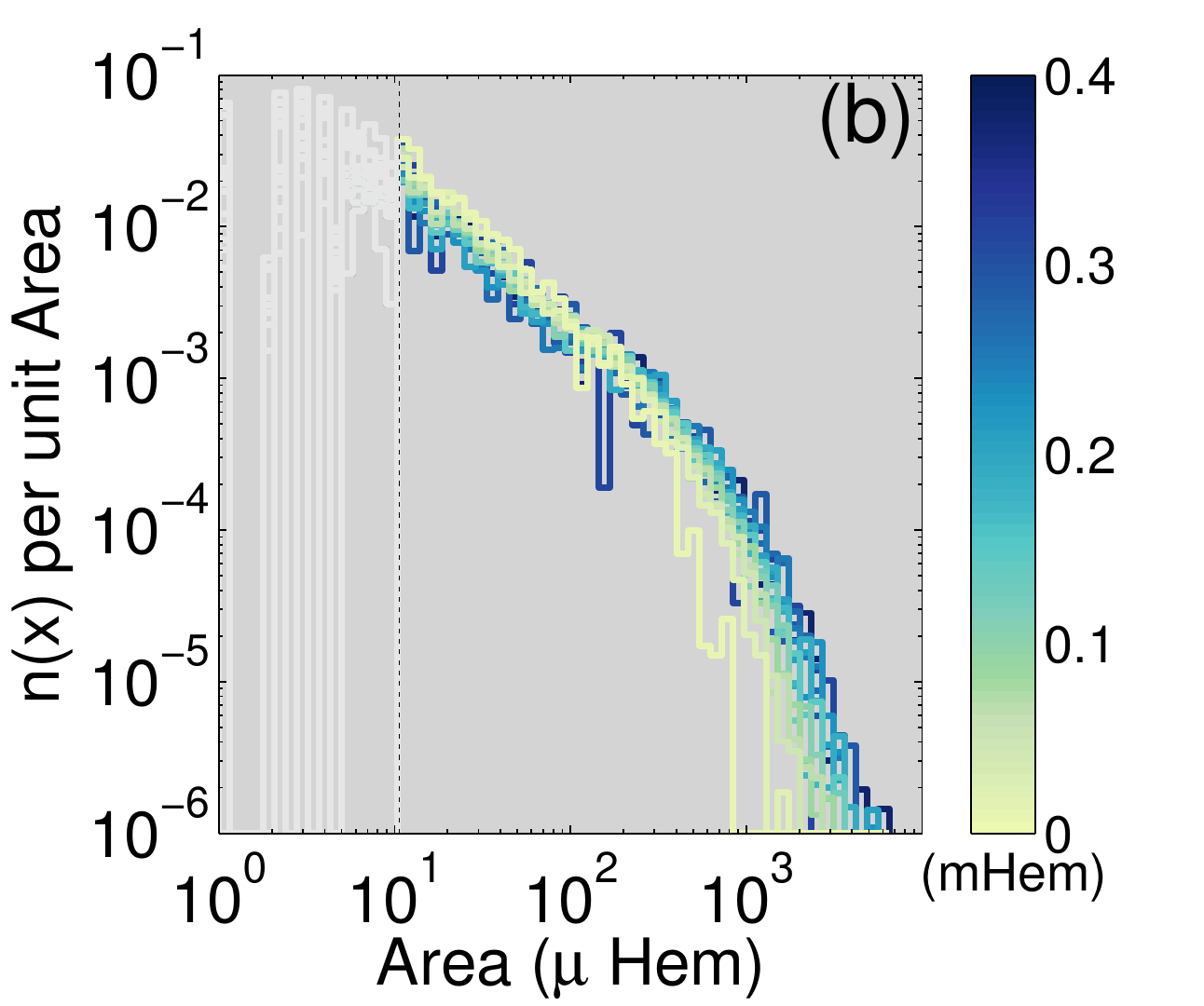}&\includegraphics[width=0.3\textwidth]{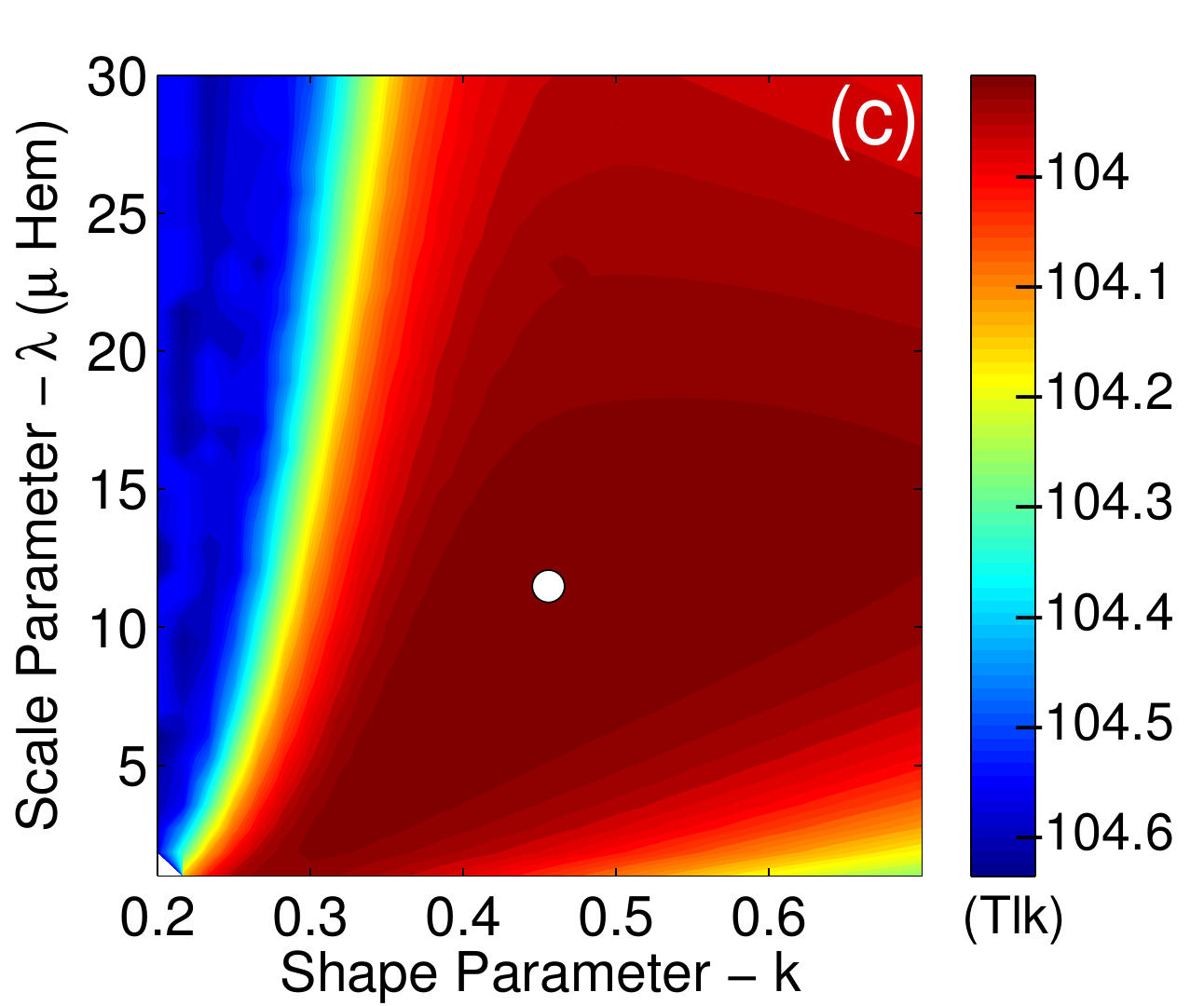}\\
  \includegraphics[width=0.3\textwidth]{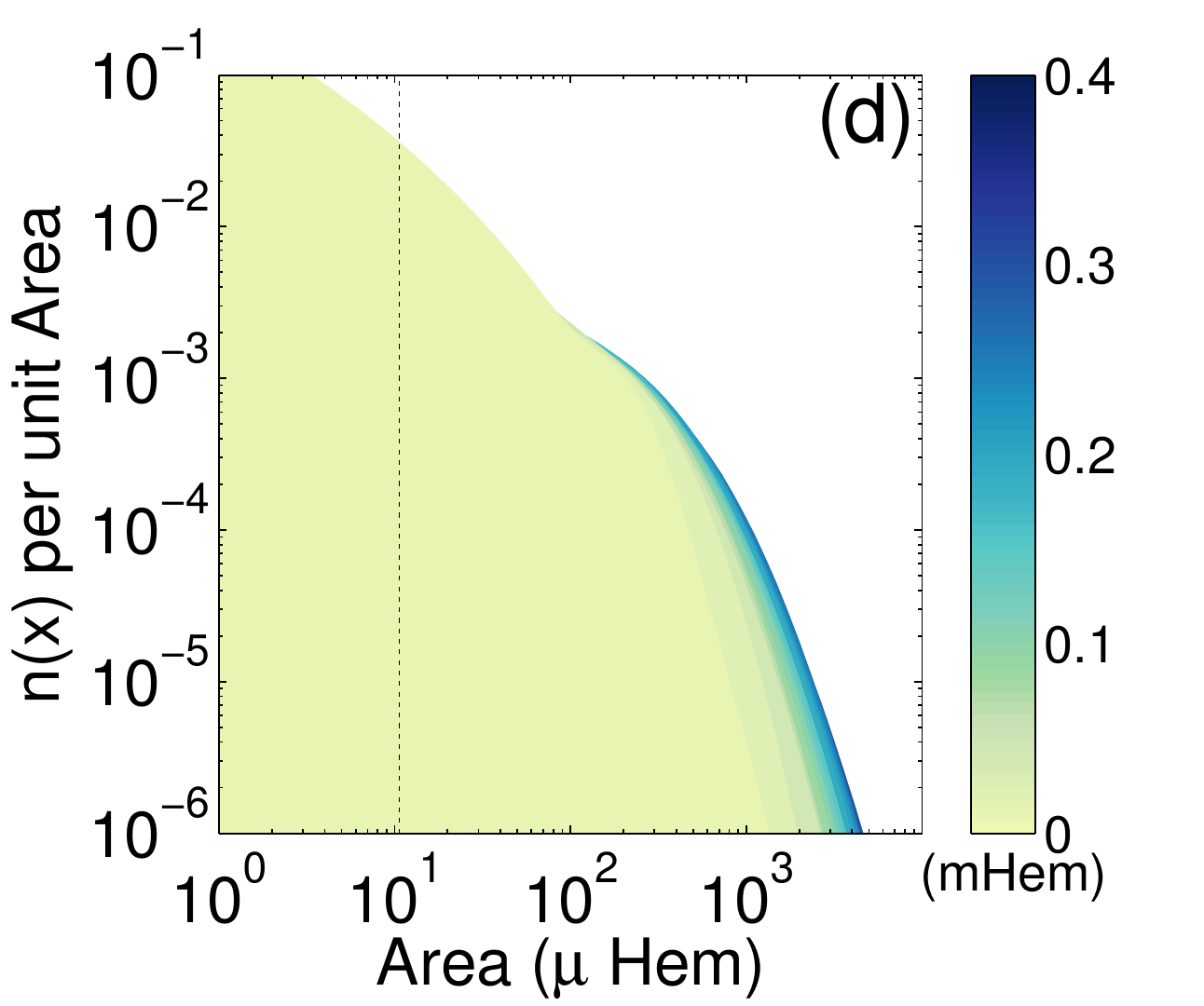}&\includegraphics[width=0.3\textwidth]{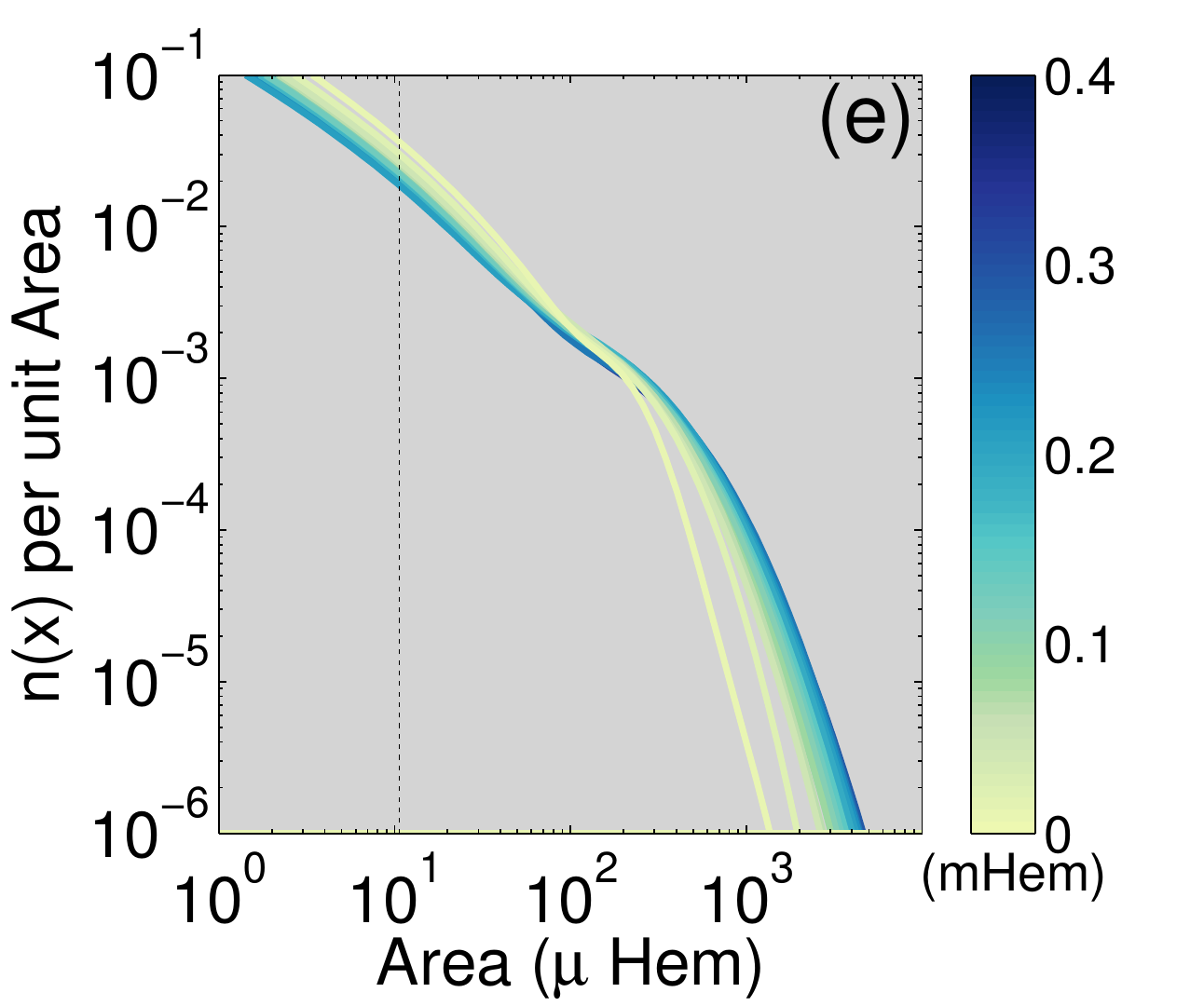}&\\
  \includegraphics[width=0.3\textwidth]{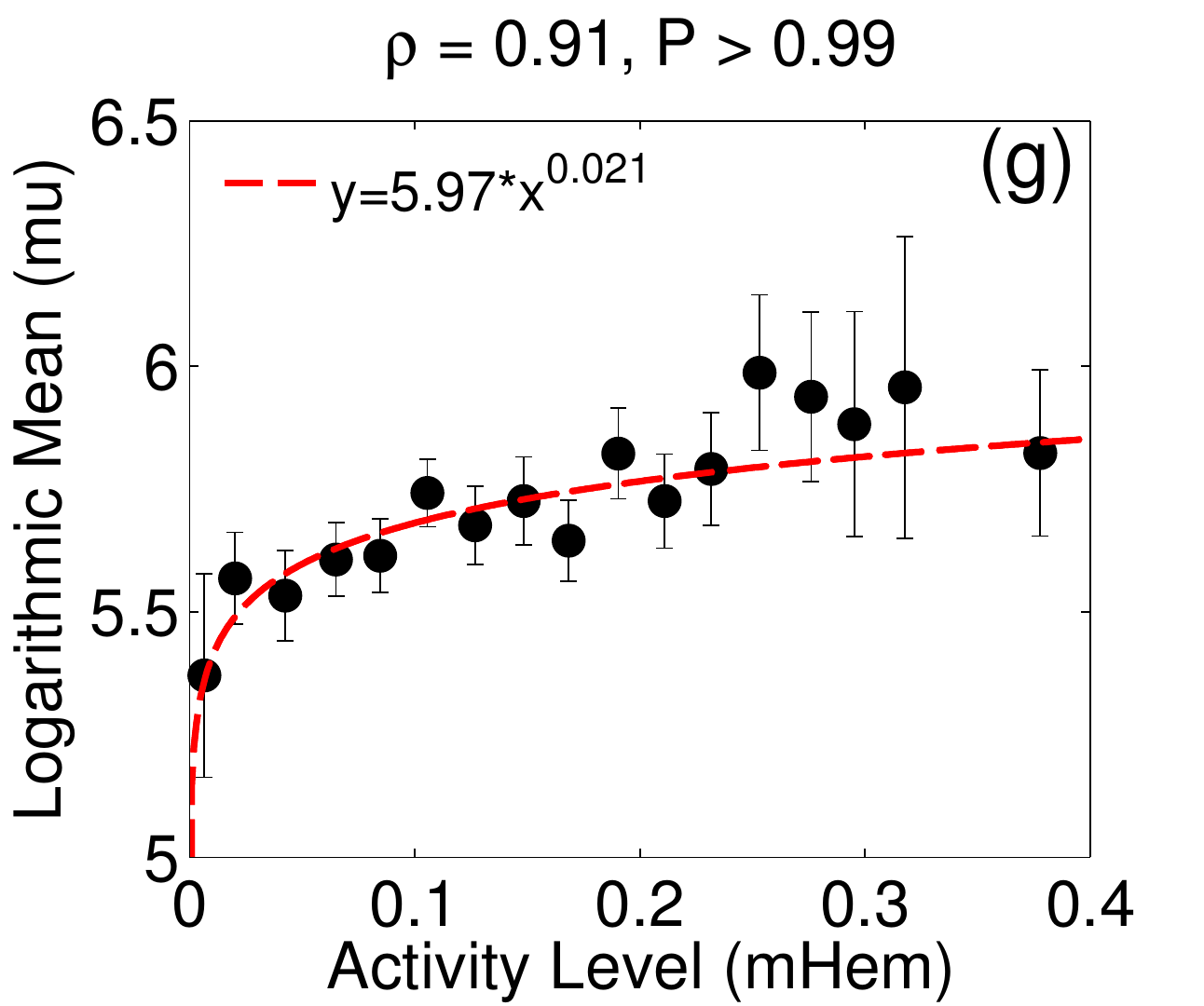}&\includegraphics[width=0.3\textwidth]{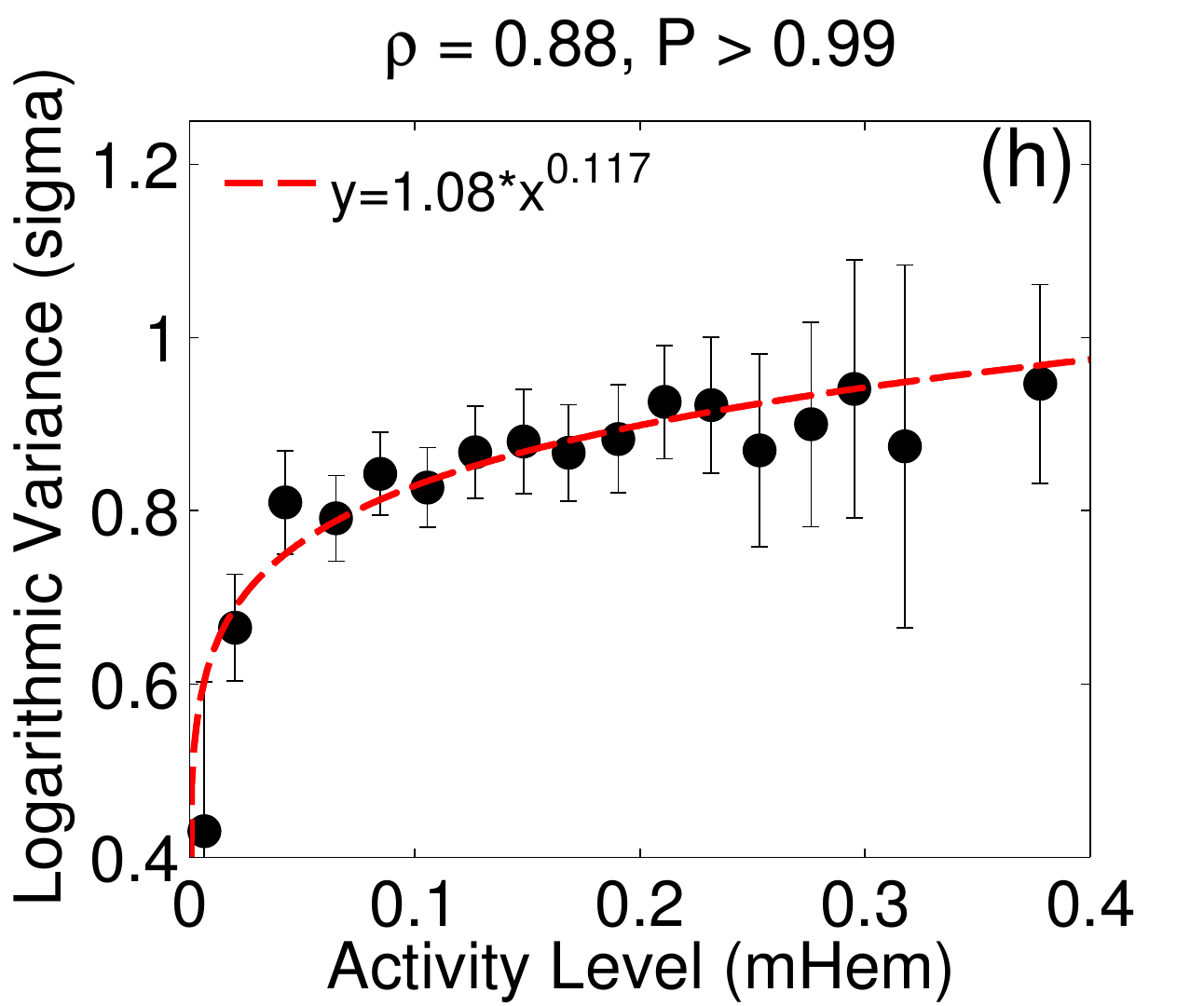}&\includegraphics[width=0.3\textwidth]{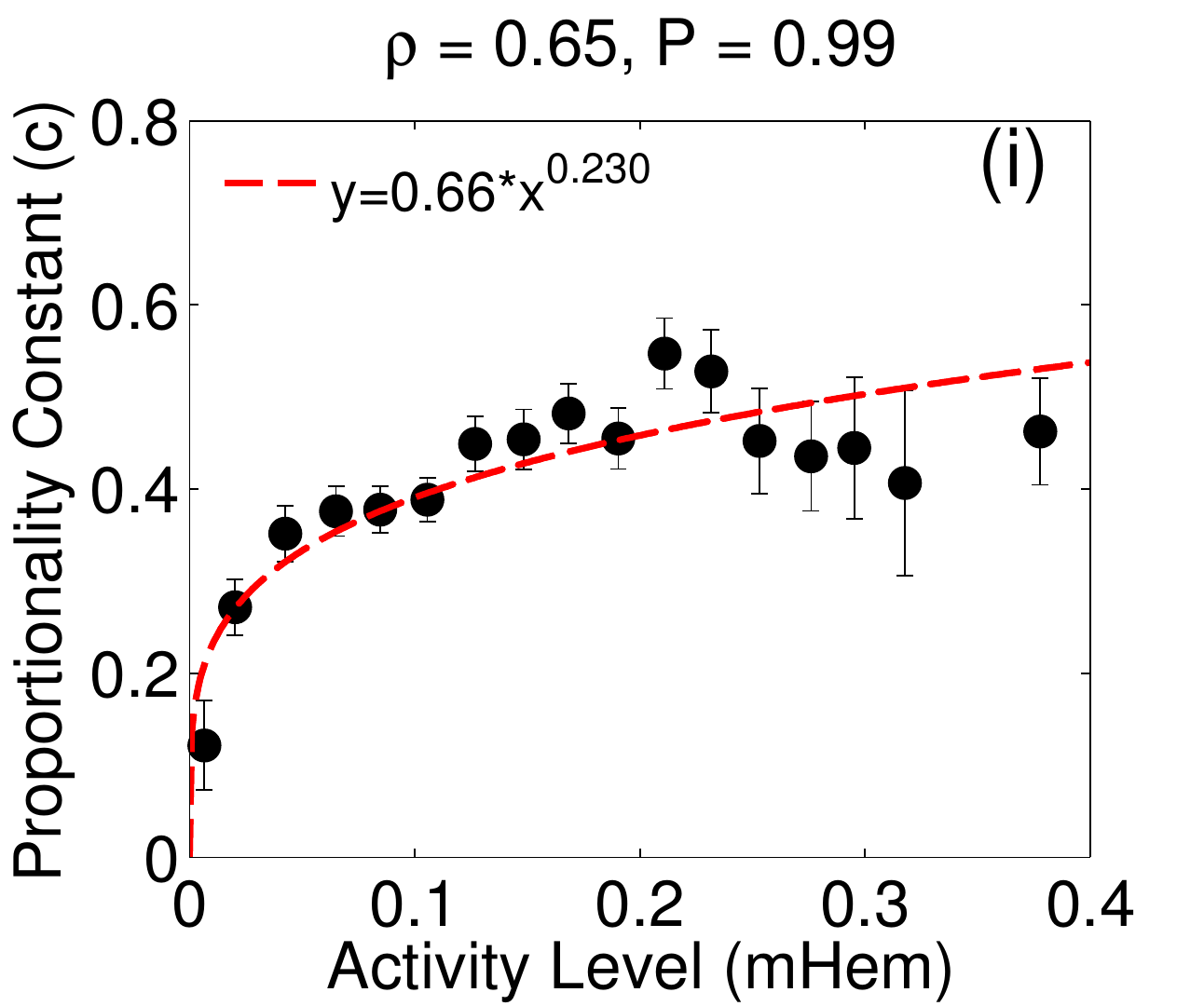}\\
\end{tabular}
\end{center}
\caption{(a), (b) Empirical size distribution associated with different activity levels. (d), (e) Composite PDFs fitted to data binned according to activity level using the same Weibull parameters for all activity levels. Vertical dashed lines mark the limit below which data are not included in the fits.  (c) Total average log-likelihood (Tlk) for all activity levels as a function of the Weibull parameters $k$ and $\lambda$ (see Equation (\ref{Eq_TLL})).  The optimum values that maximize Tlk simultaneously for all activity level bins are $k_{\text{best}}=0.46$ and $\lambda_{\text{best}}=11.49\mu$Hem.  Panels (g)-(i) show the relationship between the remaining parameters in the composite PDF and activity level.  Error bars indicate the 95\% confidence intervals of each value.  The Spearman's rank correlation coefficient ($\rho$) and its confidence level ($P$) are included as the title of each of these panels.  Fits to the relationships between these parameters and activity level are shown as red dashed lines. The analytical expression for each fit is included in the legend of each panel.}\label{Fig_Fits2}
\end{figure*}

In order to evaluate the relationship between the different fitting parameters of the composite PDF, we use Spearman's rank correlation coefficient \citep[$\rho;$][]{spearman1904}, which assesses how well the relationship between two variables can be described using a monotonic function.  The results are displayed in Figures \ref{Fig_Fits1}(c), (f)-(i).  We find no correlation between activity level and the parameters characterizing the Weibull component of the composite PDF -- with $\rho=0.0$ and $\rho=-0.19$ for the factor ($\lambda$) and shape parameter ($k$), respectively.  On the other hand, the parameters that characterize the log-normal component are found to be correlated with activity level with a high degree of statistical significance (above 98\%) -- with $\rho=0.66$ and $\rho=0.57$ for the logarithmic mean ($\mu$) and the logarithmic variance ($\sigma$) respectively.  We find a moderate correlation between the proportionality constant that blends these distributions together ($c$) with $\rho=0.37$.

\section{Using Common Weibull Parameters for All Activity Levels}\label{Sec_Dis_Con}

The apparent independence between activity level and the parameters characterizing the Weibull component of the composite PDF is in agreement with the results of \cite{hagenaar-etal2003,hagenaar-etal2008} who found essentially no dependence between the distribution of ephemeral regions and the solar cycle.  Furthermore, considering the values that these parameters assume, and the large width of their 95\% confidence intervals (see Figures \ref{Fig_Fits1}(c) and (f)), it is clear that leaving them unconstrained is being used by our fitting algorithm to over-fit the data.

To address this, we re-fit our data with the additional constraint that the parameters characterizing the Weibull component must be the same for all activity levels.  We do this by maximizing total average log-likelihood ($\operatorname{Tlk}$) for all activity levels:
\begin{equation}\label{Eq_TLL}
  \operatorname{Tlk}(k,\lambda) = \sum_{j=1}^{N_{\text{bins}}}\frac{1}{n_j}\sum_{i=1}^{n_j} \log(f(D_i^j;k,\lambda,\mu_j,\sigma_j, c_j))),
\end{equation}
where $f(x;k,\lambda,\mu,\sigma, c)$ is our composite PDF function (see Equation (\ref{Eq_Mix})); the index $j$ denotes each activity level bin, the index $i$ denotes each data point in a bin; $k$ and $\lambda$ (the Weibull parameters) are free to vary but must be the same for all activity level bins; and $\mu_j$, $\sigma_j$, and $c_j$ (the log-normal parameters and the constant of proportionality) are allowed be different for different bins.

As can be seen in Figure \ref{Fig_Fits2}(c),  $\operatorname{Tlk}$ has a single global maximum located at $k_{\text{best}}=0.46$ and $\lambda_{\text{best}}=11.49\mu$Hem.  These values are well within the 95\% confidence intervals previously found for $k$ and $\lambda$ in both the unconstrained fit (see Figures \ref{Fig_Fits1}(c) and (f)), and the fit to the unbinned RGO/KMAS Set (see Table \ref{Tab_Mix}).

After forcing $k$ and $\lambda$ to have the same value for all activity level bins, there is a remarkable tightening of the relationship between activity level and the remaining PDF parameters ($\mu$, $\sigma$, and $c$, which can be seen both qualitatively and as a significant improvement in the Spearman's rank correlation coefficients.

We perform a $\chi^2$ fit to this dependence using power functions (see Figures \ref{Fig_Fits2}(g)-(i) for fitting values), finding a reduced $\chi^2$ lower than unity in all cases.  Although in this work we fit these dependencies using power functions, due to their simplicity there are several functional forms that would fit the scatter plots equally well within the 95\% confidence intervals (for example logarithmic and exponential forms). The true characterization of these dependencies would involve a large amount of tests that is beyond the scope of this paper and will be performed in a later work.

\begin{table*}[ht!]
\begin{center}
\begin{tabular*}{1\textwidth}{@{\extracolsep{\fill}}  l c c c c c}
\multicolumn{5}{c}{\textbf{Quantification of Distribution Performance Using AIC}}\\
\toprule
 Fit Characteristics & Description & Log-Likelihood & Degrees of Freedom & $\operatorname{\Delta^{AIC}_j}$ &   Aw\\
 \midrule
 No dependence on activity level         & Section \ref{Sec_Same_Dis} & -2.097$\times$10$^5$ & 5  & 3,779 & $<$0.001\\
 Unconstrained, binned by activity level & Section \ref{Sec_Dis_Free} & -2.093$\times$10$^5$ & 85 & 3,063 & $<$0.001\\
 Constrained, binned by activity level   & Section \ref{Sec_Dis_Con}  & -2.093$\times$10$^5$ & 53 & 3,033 & $<$0.001\\
 \textbf{No binning, analytical dependence on activity level} & \textbf{Equation (\ref{Eq_Final})}  & \textbf{-2.078$\times$10$^5$} & \textbf{9}  & \textbf{0}     & \textbf{$>$0.999}\\
\end{tabular*}
\end{center}
\hspace{1em}
  \caption{Comparative performance of the different ways of fitting the data presented in this paper. $\operatorname{\Delta^{AIC}_j}$ is the relative AIC difference described by Equation (\ref{Eq_AICDel}).  Aw is the Akaike weight described by Equation (\ref{Eq_AICW}). The lower $\operatorname{\Delta^{AIC}_j}$ is, the more a model is likely to be correct (quantified using Aw).  Bold text indicates the best model according to AIC.}\label{Tab_AIC}
\end{table*}

Nevertheless, using these results, one can define a PDF with constant $k$ and $\lambda$, whose properties depend on activity level through the relationships shown in Figures \ref{Fig_Fits2}(g)-(i), and in which binning by activity level is no longer necessary.  This PDF is defined as:
\begin{equation}\label{Eq_Final}
\begin{gathered}
\begin{aligned}
  f[x;k,\lambda,\mu(AL),\sigma(AL),c(&AL)] =\nonumber\\
                              & \frac{[1-c(AL)]k}{\lambda}\left(\frac{x}{\lambda}\right)^{k-1} e^{-(x/\lambda)^k}\nonumber\\
                              &  + \frac{c(AL)}{x\sigma(AL)\sqrt{2\pi}}e^{ -\frac{(\ln x-\mu(AL))^2}{2 \sigma^2(AL)} }\nonumber
\end{aligned}\\
\begin{array}{cc}
\text{\textbf{Weibull}} & \text{\textbf{Log-normal}}\\
\begin{array}{rl}
  k =&0.46 \\
  \lambda =& 11.49\mu\text{Hem}
\end{array}&
\begin{array}{rl}
  \mu(AL) =& 5.97AL^{0.021}\\
  \sigma =& 1.08AL^{0.117}
\end{array}
\end{array}\nonumber\\
\begin{array}{c}
\text{\textbf{Proportionality Constant}}\\
c(AL) = 0.66AL^{0.230}
\end{array}\nonumber
\end{gathered}
\end{equation}
where AL is the activity level in mHem at the day each sunspot group was observed.\\


\section{Quantitative Model Selection}\label{Sec_Model_Sel}

Now that we have characterized the dependence of the size-flux PDF on activity level, our task is to quantitatively identify the best of all the models that have been used so far.  These are:

\begin{enumerate}
  \item The same PDF irrespective of activity level \citep[see ][and Section \ref{Sec_Same_Dis}]{munoz-etal2015a}.
  \item A different PDF for each activity level bin (see Section \ref{Sec_Dis_Free}).
  \item A different PDF for each activity level bin, but in which the values of $k$ and $\lambda$ are forced to be equal for every bin (see Section \ref{Sec_Dis_Con}).
  \item A PDF with constant $k$ and $\lambda$, but in which $\mu$, $\sigma$, and $c$ depend on activity level through power functions (see Equation (\ref{Eq_Final})).
\end{enumerate}

For this purpose, we use the AIC (described in detail in Appendix \ref{Sec_AIC}), and the results are shown in Table \ref{Tab_AIC}.  A comparison between log-likelihood and dof (columns 3 and 4, respectively), shows that log-likelihood is the main factor determining AIC (not the dof; see Equation (\ref{Eq_AIC})).  The reason is that our data set has significantly more data points than the dof in each of our models.

As expected, the worst fit corresponds to the PDF that does not depend on activity level (defined in Section \ref{Sec_Same_Dis}).  This is followed by both our constrained and unconstrained fits binned by activity level (defined in Sections \ref{Sec_Dis_Free} and \ref{Sec_Dis_Con}, respectively).  We find, with a very high degree of statistical significance, that the best model to fit our data is the unbinned PDF whose parameters depend analytically on activity level (defined by Equation (\ref{Eq_Final})).  It is important to highlight that AIC works only as a relative estimate.  This means that it cannot tell us whether Equation (\ref{Eq_Final}) characterizes the true mechanism that generated the observed data.  Instead, it allows us to rule out all the models proposed in this work (with near certainty) in favor of the model described by Equation (\ref{Eq_Final}).  Taken together with the results of \cite{munoz-etal2015a} who fitted five more models to similar data (including the model described in Section \ref{Sec_Same_Dis}), we can take advantage of the relative nature of AIC to rule those models out as well.

\begin{figure*}[ht!]
\begin{center}
  \includegraphics[width=0.75\textwidth]{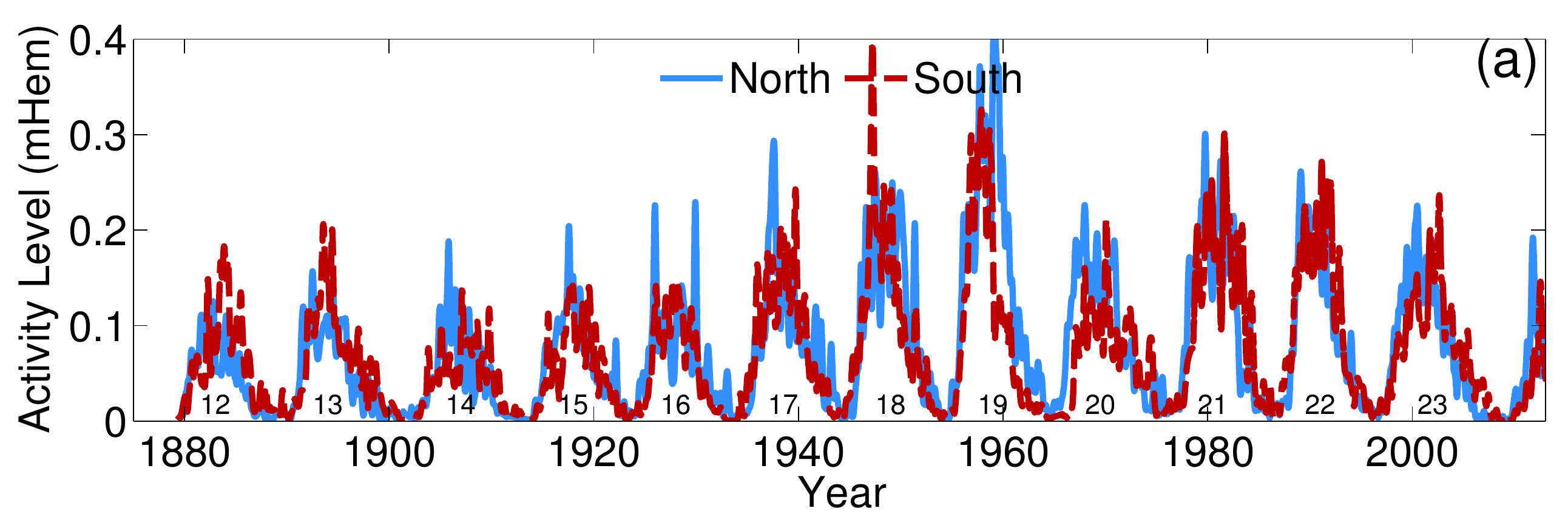}\\
  \includegraphics[width=0.75\textwidth]{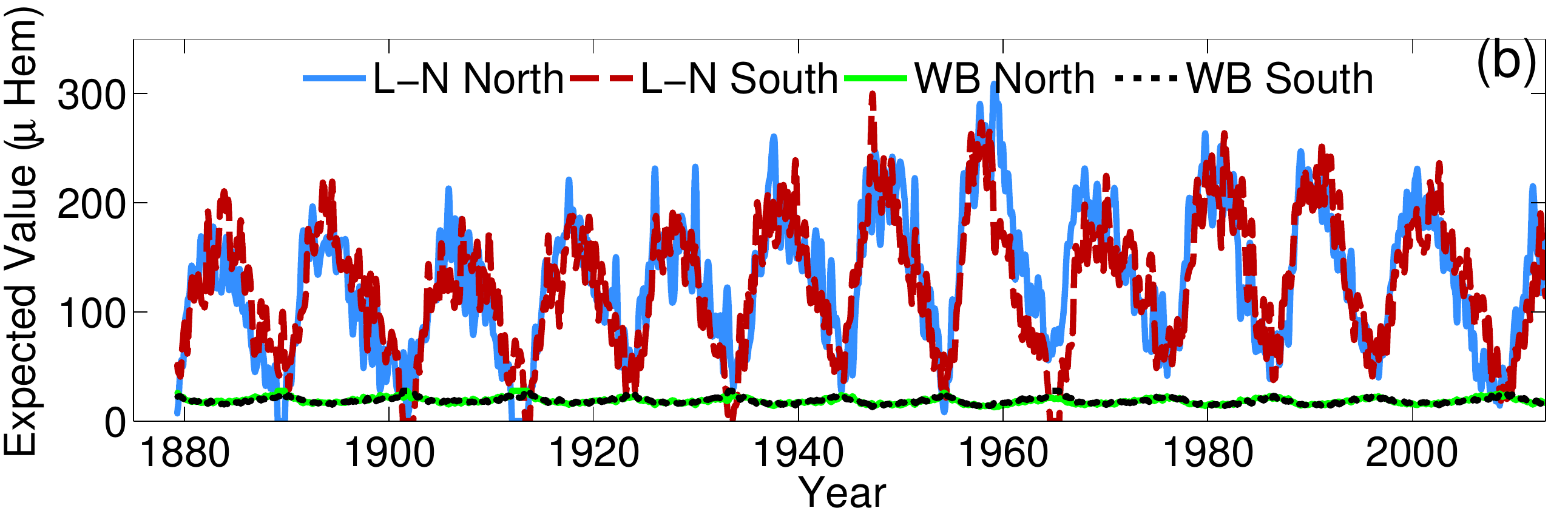}\\
  \includegraphics[width=0.75\textwidth]{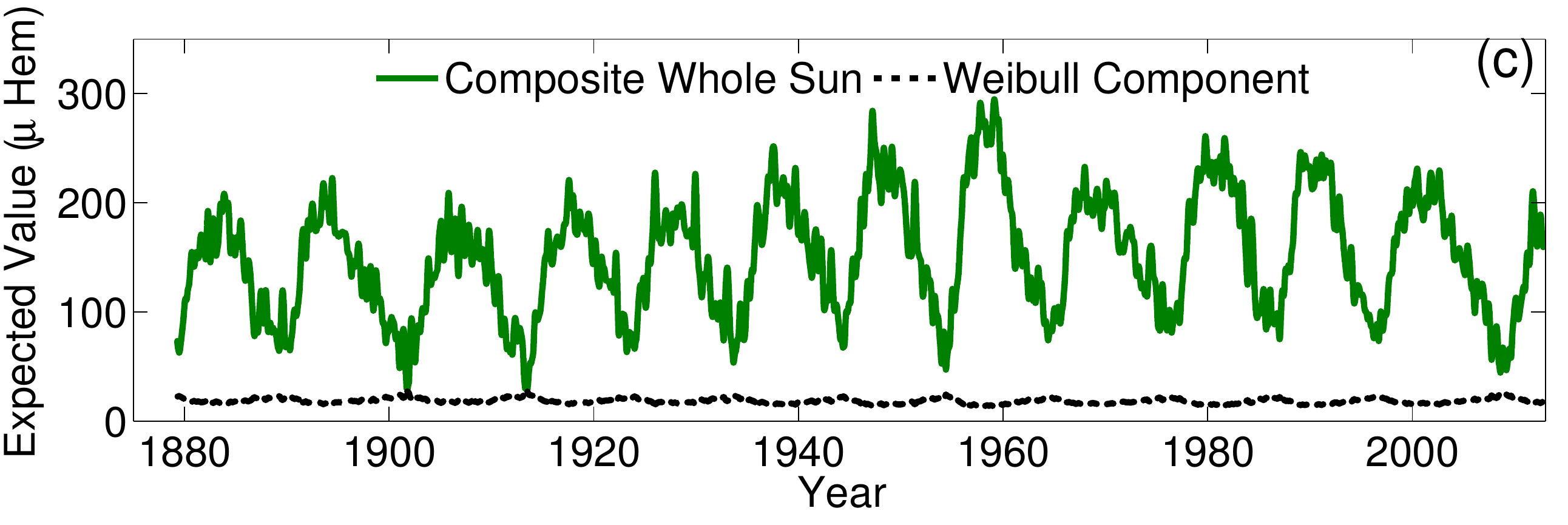}\\
\end{center}
\caption{(a) Total daily sunspot area smoothed using a six-month Gaussian filter; this is the quantity we use to define activity level. (b) Log-normal and Weibull contributions to the expected sunspot group area.  Log-normal components are calculated separately for each hemisphere; the Weibull component is shown using a black dotted line.  For both panels, the northern (southern) hemisphere is denoted by a solid light blue (dashed dark red) line. (c) Expected value of sunspot group area for the whole Sun (solid green line). For reference, the Weibull component is shown as well.}\label{Fig_Time_Dep}
\end{figure*}

\section{The Solar Magnetic Floor and the Time Dependence of Sunspot Properties}\label{Sec_Time_Dep}

As was shown in Section \ref{Sec_Same_Dis}, we find the Weibull component of the composite PDF to be independent of activity level.  This result is in agreement with other studies arguing in favor of a magnetic baseline \citep{svalgaard-cliver2007,schrijver-etal2011,cliver2012}, since it suggests that small magnetic structures in the photosphere indeed arise from a cycle-independent process.  However, the striking connection between the size-flux PDF and activity level (demonstrated in Section \ref{Sec_Dis_Con}), allows us to do more than that; it allows us make a quantitative comparison of the depth of each minimum during the last 130 yr (12 solar cycles).  For this purpose, we calculate the expected value of sunspot group area (i.e., the average size of magnetic structures) as a function of activity level:
\begin{equation}\label{Eq_Exp_Val}
\begin{gathered}
  \begin{aligned}
  \mathbb{E}[f(AL)]  = {} & \int_{0}^{\infty} x f[x;k,\lambda,\mu(AL),\sigma(AL), c(AL)]dx\\
                    = {} & [1-c(AL)] \lambda \Gamma \left(1+\frac{1}{k} \right)\\
                         & + c(AL)e^{\mu(AL)+\frac{\sigma^2(AL)}{2}},
  \end{aligned}
\end{gathered}
\end{equation}
where $\Gamma$ is the gamma function.  This quantity can be used as a thermometer for solar magnetism, as it tells us the typical magnetic structure size that we can expect to see as a function of time.

As it happens with the composite PDF, the expected value of the sunspot group area is also a linear combination of Weibull (second line, Equation (\ref{Eq_Exp_Val})) and log-normal (third line, Equation (\ref{Eq_Exp_Val})) components. Furthermore, almost all of the time dependence of the expected value can be attributed to the log-normal component.  The consequence is that the magnetic baseline is defined by the Weibull component and is only truly visible in those times in which the emergence of cycle-related BMRs shuts down.

Figure \ref{Fig_Time_Dep}(b) shows the time evolution of the log-normal component of the expected sunspot area and how it contrasts with its Weibull counterpart.  Note that there is a weak modulation of the Weibull contribution to the expected value due to the fact that the proportionality constant ($c$) by definition depends on activity level.  We find that the log-normal contribution drops below its Weibull counterpart in most hemispheric minima.  Using a notation where $n.5$ corresponds to the minimum between cycle n and cycle n+1, the exceptions are 15.5N\&S, 17N\&S, 19.5N, 20.5N\&S, 21.5N\&S, and 22.5N\&S (note that most exceptions occur during the space age).  This means that, from a hemispheric point of view, we have been able to observe that magnetic baseline.  However, the results are different from a whole Sun point of view.  We calculate the whole Sun expected value as a weighted average between the expected value of the northern and southern hemispheres.  Due to its strong correlation with sunspot numbers, for simplicity we use activity level as our weighting coefficient:
\begin{equation}\label{Eq_ExWS}
   \mathbb{E}_{WS} = \frac{AL_N \mathbb{E}_{N}+AL_S \mathbb{E}_{S}}{AL_N+AL_S},
\end{equation}
where $\mathbb{E}_{WS}$, $\mathbb{E}_{N}$, and $\mathbb{E}_{S}$ are the whole Sun, northern, and southern expected values, respectively, and AL$_N$ (AL$_S$) is the activity level in the northern (southern) hemisphere.

Figure \ref{Fig_Time_Dep}(c) shows the time evolution of the expected value for the whole Sun (including both log-normal and Weibull components) and the Weibull baseline for comparison. The story is quite different; there have been only two solar minima during the last 12 cycles in which observation of the baseline magnetism has been possible in both hemispheres simultaneously (i.e., for the whole Sun; 13.5 and 14.5).  Apart from those, hemispheric asymmetries have conspired to raise the whole Sun level above the baseline magnetism.

One of the striking features evidenced in Figure \ref{Fig_Time_Dep}(c) is how different the minimum of cycle 23 is when compared to other minima during the space age (19.5 through 23.5).  It is no wonder why it seemed so unusual to us.  This has important implications because the properties of the heliosphere, its current sheet, and the background solar wind are strongly determined at a global level.  The fact that the minimum of cycle 23 had a very asymmetric current sheet is evidence that solar magnetism was not in its baseline state. Based on this result, we infer that even though each hemisphere did reach the magnetic baseline, the minimum of cycle 23 was not as deep as it could have been.

\section{Small-Scale Versus Global Dynamo?}\label{Sec_Dynamo}

In our previous paper \citep{munoz-etal2015a}, we proposed that the existence of two populations of sunspot groups originated from separate contributions by the small-scale and global components of the dynamo.  However, with the evolution of our understanding we feel that further clarification is necessary.  We find strong evidence that the size-flux distribution has two components that are discriminated by size, and only the properties of the larger structures are modulated by the solar cycle.  However, in spite of the fact that the properties of small pores are independent of the solar cycle, their latitude of appearance is still modulated by active latitudes (as can be seen in Figure (\ref{Fig_Bfly_Bin})).  This means that there must still be a connection between these structures and the cycle itself and that their formation cannot be attributed solely to the small-scale component of the dynamo (otherwise they would be observable all throughout the photosphere).

A possible explanation is that these small structures arise from the re-processing of the decaying magnetic field of their large-scale counterparts and thus their relative numbers are governed by the properties of surface convection (and only loosely by the amount of available decaying field). In this case, the magnetic baseline found in the previous section is contingent on the emergence of a minimum amount of large-scale structures and cannot be taken as a hard lower limit for grand minima like the Maunder minimum. Unfortunately, it is difficult to make a further assessment of this connection without resorting to magnetic data.  For this reason, we will look at this issue in more detail in future work involving magnetic structure catalogs compiled using SOHO/MDI and SDO/HMI data.

\section{Summary and Concluding Remarks}\label{Sec_Conclusions}

In this work we have introduced a new way of binning sunspot group and BMR data with the purpose of better understanding the impact of the solar cycle on sunspot and BMR properties and how this defines the characteristics of the extended minimum of cycle 23.  This approach hinges critically on our current understanding of BMRs as the photospheric manifestation of emergent buoyant flux tubes arising from a large-scale underlying toroidal field.  In particular, we assume that from the point of view of each active region, the solar cycle can be approximated as a quasi-static process.  This means that the properties of sunspots and BMRs are completely determined by the strength of the underlying toroidal field and have no additional long-term dependencies.  In other words, we are assuming that the statistical properties of sunspots and BMRs do not depend on cycle phase (rising versus declining; maximum versus minimum), but on how strong the cycle is at each particular moment (something we refer to as activity level).

In this work we build upon the results of \cite{munoz-etal2015a}, who found after analyzing 11 different databases, that the solar size-flux distribution is better characterized by a linear combination of Weibull and log-normal distributions -- where a pure Weibull (log-normal) characterizes the distribution of structures with fluxes below (above) $10^{21}$Mx ($10^{22}$Mx).   After binning our data according to activity level, we fit this composite distribution to each separate bin and look at the dependence of each parameter on activity level.

We find that the parameters that characterize the Weibull component have no dependence on activity level.  This is in agreement with the results of \cite{hagenaar-etal2003,hagenaar-etal2008} who found essentially no dependence between the distribution of ephemeral regions (below $10^{20}$Mx) and the solar cycle.  We propose that the structures characterized by the Weibull component are what give the Sun a magnetic baseline.

In stark contrast to the Weibull component, we find a clear dependence between activity level and the parameters that characterize the log-normal component of the size-flux distribution (which we fit using power functions).  This supports the interpretation of \cite{munoz-etal2015a}, who proposed that the log-normal component is directly connected to the global component of the dynamo (and the generation of bipolar active regions).

By taking advantage of our analytical characterization of the size-flux distribution and its dependence on activity level, we study the relative contribution of each component (small-scale versus large-scale) to solar magnetism.  In order to do this, we calculate the expected value of sunspot group areas and study its evolution with time.  We find that from a hemispheric point of view, almost every solar minimum (during the last 12 cycles) reaches a point where the only contribution to magnetism comes from the small-scale component.  However, due to asymmetries in cycle phase, this state is very rarely reached by both hemispheres at the same time (according to our data, only during the minima of cycles 13 and 14). There is no question that the extended minimum of cycle 23 is deeper than any other minimum of the space age.  However, based on our results, we infer that even though each hemisphere did reach the magnetic baseline, from a heliospheric point of view the minimum of cycle 23 was not as deep as it could possibly be.

\acknowledgements

\section{Acknowledgements}

We thank Karel Schrijver, Mar\'ia Navas Moreno, and an anonymous referee for useful discussions and suggestions.  This research was supported by the NASA Living With a Star Jack Eddy Postdoctoral Fellowship Program, administered by the UCAR Visiting Scientist Programs, contract SP02H1701R from Lockheed-Martin to the Smithsonian Astrophysical Observatory, and the CfA Solar Physics REU program, NSF grant number AGS-1263241.  Andr\'es Mu\~noz-Jaramillo is very grateful to George Fisher and Stuart Bale for their support at the University of California, Berkeley, and Phil Scherrer for his support at Stanford University.  The National Solar Observatory (NSO) is operated by the Association of Universities for Research in Astronomy, AURA Inc. under cooperative agreement with the National Science Foundation (NSF).

\appendix

\section{Maximum Likelihood Estimation}\label{Sec_MLE}

The idea behind MLE, is to find the set of parameters that maximizes the likelihood of a statistical model $M$ given the observed data $D = \{ D_1, D_2, ... , D_n \}$ by maximizing the likelihood ($\operatorname{L}$) function:
\begin{equation}\label{Eq_L}
  \operatorname{L}(M)\propto \operatorname{pr}(D|M) = \prod_{i=1}^n \operatorname{pr}(D_i|M).
\end{equation}
This process of maximization is typically performed by first taking the logarithm of both sides of Equation (\ref{Eq_L}), and maximizing the resulting log-likelihood ($\operatorname{lk}$) function:
\begin{equation}\label{Eq_LL}
  \operatorname{lk}(M) = \sum_{i=1}^n \log(\operatorname{pr}(D_i|M)).
\end{equation}

\section{Akaike's Information Criterion}\label{Sec_AIC}

The AIC for a model $M_j$ is defined as:
\begin{equation}\label{Eq_AIC}
  \operatorname{AIC_j} = - 2 \operatorname{lk}(M_j) - 2 n_j,
\end{equation}
where $\operatorname{lk}(M_j)$  is the log-likelihood of model $M_j$ (see Equation (\ref{Eq_LL})) and $n_j$ is the number of parameters of model $j$.  The model with the minimum AIC is chosen as the best.  In a sense, by minimizing the AIC one is looking for the model with the largest log-likelihood.  However, log-likelihood alone is not sufficient to discriminate between models because it is biased as an estimation of the model selection target.  This bias was found by \cite{akaike1983} to be approximately equal to each model's number of parameters ($n$) and thus the presence of the second term in Equation (\ref{Eq_AIC}).  The significance of the AIC is strongly dependent on an appropriate choice of models.  Applying the AIC to a set of very poor models will always select one estimated to be the best (even though that model may still be poor in an absolute sense).

The relative nature of the AIC is better represented by calculating the relative AIC differences:
\begin{equation}\label{Eq_AICDel}
  \operatorname{\Delta^{AIC}_j} = \operatorname{AIC_j} - \min(\operatorname{AIC}).
\end{equation}
This in turn can be used to estimate the likelihood of a model given the data:
\begin{equation}\label{Eq_AICL}
  \mathcal{L}(M_j|D) \propto \exp\left(-\frac{ \operatorname{\Delta^{AIC}_j}}{2}\right),
\end{equation}
and use it to calculate the Akaike weights:
\begin{equation}\label{Eq_AICW}
  Aw_j = \frac{\exp\left(-\frac{ \operatorname{\Delta^{AIC}_j}}{2}\right)}{\sum_{k=1}^K \exp\left(-\frac{ \operatorname{\Delta^{AIC}_k}}{2}\right)},
\end{equation}
which are a measure of the probability that the model $M_j$ is the best model given the data.\\\\

\bibliographystyle{apj}

\begin{thebibliography}{47}
\expandafter\ifx\csname natexlab\endcsname\relax\def\natexlab#1{#1}\fi

\bibitem[{Akaike(1983)}]{akaike1983}
Akaike, H. 1983, Bulletin of the International Statistical Institute, 50, 277

\bibitem[{{Babcock}(1961)}]{babcock1961}
{Babcock}, H.~W. 1961, \apj, 133, 572

\bibitem[{{Baumann} \& {Solanki}(2005)}]{baumann-solanki2005}
{Baumann}, I. \& {Solanki}, S.~K. 2005, \aap, 443, 1061

\bibitem[{{Bogdan} {et~al.}(1988){Bogdan}, {Gilman}, {Lerche}, \&
  {Howard}}]{bogdan-etal1988}
{Bogdan}, T.~J., {Gilman}, P.~A., {Lerche}, I., \& {Howard}, R. 1988, \apj,
  327, 451

\bibitem[{Burnham \& Anderson(2002)}]{burnham-anderson2002}
Burnham, K. \& Anderson, D. 2002, Model Selection and Multimodel Inference: A
  Practical Information-Theoretic Approach (Springer)

\bibitem[{{Cameron} {et~al.}(2010){Cameron}, {Jiang}, {Schmitt}, \&
  {Sch{\"u}ssler}}]{cameron-etal2010}
{Cameron}, R.~H., {Jiang}, J., {Schmitt}, D., \& {Sch{\"u}ssler}, M. 2010,
  \apj, 719, 264

\bibitem[{{Charbonneau}(2010)}]{charbonneau2010}
{Charbonneau}, P. 2010, Living Reviews in Solar Physics, 7, 3

\bibitem[{{Cliver}(2012)}]{cliver2012}
{Cliver}, E.~W. 2012, in IAU Symposium, Vol. 286, IAU Symposium, ed. C.~H.
  {Mandrini} \& D.~F. {Webb}, 179--184

\bibitem[{{Dasi-Espuig} {et~al.}(2010){Dasi-Espuig}, {Solanki}, {Krivova},
  {Cameron}, \& {Pe{\~n}uela}}]{dasiespuig-etal2010}
{Dasi-Espuig}, M., {Solanki}, S.~K., {Krivova}, N.~A., {Cameron}, R., \&
  {Pe{\~n}uela}, T. 2010, \aap, 518, A7

\bibitem[{{Dasi-Espuig} {et~al.}(2013){Dasi-Espuig}, {Solanki}, {Krivova},
  {Cameron}, \& {Pe{\~n}uela}}]{dasiespuig-etal2013}
---. 2013, \aap, 556, C3

\bibitem[{{de Toma} {et~al.}(2013){de Toma}, {Chapman}, {Preminger}, \&
  {Cookson}}]{detoma-etal2013a}
{de Toma}, G., {Chapman}, G.~A., {Preminger}, D.~G., \& {Cookson}, A.~M. 2013,
  \apj, 770, 89

\bibitem[{{Fan}(2009)}]{fan2009}
{Fan}, Y. 2009, Living Reviews in Solar Physics, 6, 4

\bibitem[{{Hagenaar} {et~al.}(2008){Hagenaar}, {De Rosa}, \&
  {Schrijver}}]{hagenaar-etal2008}
{Hagenaar}, H.~J., {De Rosa}, M.~L., \& {Schrijver}, C.~J. 2008, \apj, 678, 541

\bibitem[{{Hagenaar} {et~al.}(2003){Hagenaar}, {Schrijver}, \&
  {Title}}]{hagenaar-etal2003}
{Hagenaar}, H.~J., {Schrijver}, C.~J., \& {Title}, A.~M. 2003, \apj, 584, 1107

\bibitem[{{Harvey} \& {Zwaan}(1993)}]{harvey-zwaan1993}
{Harvey}, K.~L. \& {Zwaan}, C. 1993, \solphys, 148, 85

\bibitem[{{Hathaway}(2010)}]{hathaway2010}
{Hathaway}, D.~H. 2010, Living Reviews in Solar Physics, 7, 1

\bibitem[{{Hathaway} \& {Choudhary}(2008)}]{hathaway-choudhary2008}
{Hathaway}, D.~H. \& {Choudhary}, D.~P. 2008, \solphys, 250, 269

\bibitem[{Hoel(1984)}]{hoel1984}
Hoel, P.~G. 1984, Introduction to mathematical statistics (Wiley New York), 1
  v.

\bibitem[{{Ivanov}(2012)}]{ivanov2012}
{Ivanov}, V.~G. 2012, Geomagnetism and Aeronomy, 52, 999

\bibitem[{{Jiang} {et~al.}(2011){Jiang}, {Cameron}, {Schmitt}, \&
  {Sch{\"u}ssler}}]{jiang-etal2011a}
{Jiang}, J., {Cameron}, R.~H., {Schmitt}, D., \& {Sch{\"u}ssler}, M. 2011,
  \aap, 528, A82

\bibitem[{{Kuklin}(1980)}]{kuklin1980}
{Kuklin}, G.~V. 1980, Bulletin of the Astronomical Institutes of
  Czechoslovakia, 31, 224

\bibitem[{{Leighton}(1964)}]{leighton1964}
{Leighton}, R.~B. 1964, \apj, 140, 1547

\bibitem[{{Mathew} {et~al.}(2007){Mathew}, {Mart{\'{\i}}nez Pillet}, {Solanki},
  \& {Krivova}}]{mathew-etal2007}
{Mathew}, S.~K., {Mart{\'{\i}}nez Pillet}, V., {Solanki}, S.~K., \& {Krivova},
  N.~A. 2007, \aap, 465, 291

\bibitem[{{McClintock} \& {Norton}(2013)}]{mcclintock-norton2013}
{McClintock}, B.~H. \& {Norton}, A.~A. 2013, \solphys, 287, 215

\bibitem[{{Meunier}(2003)}]{meunier2003}
{Meunier}, N. 2003, \aap, 405, 1107

\bibitem[{{Mu{\~n}oz-Jaramillo} {et~al.}(2013){Mu{\~n}oz-Jaramillo},
  {Dasi-Espuig}, {Balmaceda}, \& {DeLuca}}]{munoz-etal2013a}
{Mu{\~n}oz-Jaramillo}, A., {Dasi-Espuig}, M., {Balmaceda}, L.~A., \& {DeLuca},
  E.~E. 2013, \apjl, 767, L25

\bibitem[{{Mu{\~n}oz-Jaramillo} {et~al.}(2015){Mu{\~n}oz-Jaramillo},
  {Senkpeil}, {Windmueller}, {Amouzou}, {Longcope}, {Tlatov}, {Nagovitsyn},
  {Pevtsov}, {Chapman}, {Cookson}, {Yeates}, {Watson}, {Balmaceda}, {DeLuca},
  \& {Martens}}]{munoz-etal2015a}
{Mu{\~n}oz-Jaramillo}, A., {Senkpeil}, R.~R., {Windmueller}, J.~C., {Amouzou},
  E.~C., {Longcope}, D.~W., {Tlatov}, A.~G., {Nagovitsyn}, Y.~A., {Pevtsov},
  A.~A., {Chapman}, G.~A., {Cookson}, A.~M., {Yeates}, A.~R., {Watson}, F.~T.,
  {Balmaceda}, L.~A., {DeLuca}, E.~E., \& {Martens}, P.~C.~H. 2015, \apj, 800,
  48

\bibitem[{{Nagovitsyn} {et~al.}(2012){Nagovitsyn}, {Pevtsov}, \&
  {Livingston}}]{nagovitsyn-etal2012}
{Nagovitsyn}, Y.~A., {Pevtsov}, A.~A., \& {Livingston}, W.~C. 2012, \apjl, 758,
  L20

\bibitem[{{Parker}(1955)}]{parker1955a}
{Parker}, E.~N. 1955, \apj, 122, 293

\bibitem[{{Parnell}(2002)}]{parnell2002}
{Parnell}, C.~E. 2002, \mnras, 335, 389

\bibitem[{{Parnell} {et~al.}(2009){Parnell}, {DeForest}, {Hagenaar},
  {Johnston}, {Lamb}, \& {Welsch}}]{parnell-etal2009}
{Parnell}, C.~E., {DeForest}, C.~E., {Hagenaar}, H.~J., {Johnston}, B.~A.,
  {Lamb}, D.~A., \& {Welsch}, B.~T. 2009, \apj, 698, 75

\bibitem[{{Penn} \& {Livingston}(2006)}]{penn-livingston2006}
{Penn}, M.~J. \& {Livingston}, W. 2006, \apjl, 649, L45

\bibitem[{{Penn} \& {Livingston}(2011)}]{penn-livingston2011}
{Penn}, M.~J. \& {Livingston}, W. 2011, in IAU Symposium, Vol. 273, IAU
  Symposium, 126--133

\bibitem[{{Pevtsov} {et~al.}(2014){Pevtsov}, {Bertello}, {Tlatov}, {Kilcik},
  {Nagovitsyn}, \& {Cliver}}]{pevtsov-etal2014}
{Pevtsov}, A.~A., {Bertello}, L., {Tlatov}, A.~G., {Kilcik}, A., {Nagovitsyn},
  Y.~A., \& {Cliver}, E.~W. 2014, \solphys, 289, 593

\bibitem[{{Schad} \& {Penn}(2010)}]{schad-penn2010}
{Schad}, T.~A. \& {Penn}, M.~J. 2010, \solphys, 262, 19

\bibitem[{{Schrijver} {et~al.}(2011){Schrijver}, {Livingston}, {Woods}, \&
  {Mewaldt}}]{schrijver-etal2011}
{Schrijver}, C.~J., {Livingston}, W.~C., {Woods}, T.~N., \& {Mewaldt}, R.~A.
  2011, \grl, 38, 6701

\bibitem[{{Schrijver} {et~al.}(1997){Schrijver}, {Title}, {van Ballegooijen},
  {Hagenaar}, \& {Shine}}]{schrijver-etal1997}
{Schrijver}, C.~J., {Title}, A.~M., {van Ballegooijen}, A.~A., {Hagenaar},
  H.~J., \& {Shine}, R.~A. 1997, \apj, 487, 424

\bibitem[{{Spearman}(1904)}]{spearman1904}
{Spearman}, C. 1904, Am. J. Phys, 15, 72

\bibitem[{{Svalgaard} \& {Cliver}(2007)}]{svalgaard-cliver2007}
{Svalgaard}, L. \& {Cliver}, E.~W. 2007, \apjl, 661, L203

\bibitem[{{Tang} {et~al.}(1984){Tang}, {Howard}, \& {Adkins}}]{tang-etal1984}
{Tang}, F., {Howard}, R., \& {Adkins}, J.~M. 1984, \solphys, 91, 75

\bibitem[{{Wang} \& {Sheeley}(1989)}]{wang-sheeley1989}
{Wang}, Y. \& {Sheeley}, Jr., N.~R. 1989, \solphys, 124, 81

\bibitem[{{Wang} \& {Sheeley}(2013)}]{wang-sheeley2013}
{Wang}, Y.-M. \& {Sheeley}, Jr., N.~R. 2013, \apj, 764, 90

\bibitem[{{Watson} {et~al.}(2011){Watson}, {Fletcher}, \&
  {Marshall}}]{watson-fletcher-marshall2011}
{Watson}, F.~T., {Fletcher}, L., \& {Marshall}, S. 2011, \aap, 533, A14

\bibitem[{{Weber} {et~al.}(2011){Weber}, {Fan}, \& {Miesch}}]{weber-etal2011}
{Weber}, M.~A., {Fan}, Y., \& {Miesch}, M.~S. 2011, \apj, 741, 11

\bibitem[{{Zhang} {et~al.}(2010){Zhang}, {Wang}, \& {Liu}}]{zhang-wang-liu2010}
{Zhang}, J., {Wang}, Y., \& {Liu}, Y. 2010, \apj, 723, 1006

\bibitem[{{Zharkov} {et~al.}(2005){Zharkov}, {Zharkova}, \&
  {Ipson}}]{zharkov-zharkova-ipson2005}
{Zharkov}, S., {Zharkova}, V.~V., \& {Ipson}, S.~S. 2005, \solphys, 228, 377

\bibitem[{{Zharkova} \& {Zharkov}(2008)}]{zharkova-zharkov2008}
{Zharkova}, V.~V. \& {Zharkov}, S.~I. 2008, Advances in Space Research, 41, 881

\end{thebibliography}

\end{document}